%% file: ms.tex
\documentclass{emulateapj}

\usepackage{graphicx}
\usepackage{epstopdf}
\usepackage{amsmath}
\usepackage{hyperref}
\usepackage{cleveref}

\def\simlt{\mathrel{\hbox{\rlap{\hbox{\lower4pt\hbox{$\sim$}}}\hbox{$<$}}}}
\def\simgt{\mathrel{\hbox{\rlap{\hbox{\lower4pt\hbox{$\sim$}}}\hbox{$>$}}}}

\def\ale{\mathrel{\hbox{\rlap{\hbox{\lower4pt\hbox{$\sim$}}}\hbox{$<$}}}}
\def\age{\mathrel{\hbox{\rlap{\hbox{\lower4pt\hbox{$\sim$}}}\hbox{$>$}}}}

\def\gs{\mathrel{\raise0.35ex\hbox{$\scriptstyle >$}\kern-0.6em
\lower0.40ex\hbox{{$\scriptstyle \sim$}}}}
\def\ls{\mathrel{\raise0.35ex\hbox{$\scriptstyle <$}\kern-0.6em
\lower0.40ex\hbox{{$\scriptstyle \sim$}}}}

\def\spose#1{\hbox to 0pt{#1\hss}}
\def\simlt{\mathrel{\spose{\lower 3pt\hbox{$\mathchar"218$}}
     \raise 2.0pt\hbox{$\mathchar"13C$}}}
\def\simgt{\mathrel{\spose{\lower 3pt\hbox{$\mathchar"218$}}
     \raise 2.0pt\hbox{$\mathchar"13E$}}}

\slugcomment{\it To be submitted to the Astrophysical Journal}

\shorttitle{Late-time H$\alpha$ Emission from SLSNe-I }
\shortauthors{Yan et al.}

\begin{document}

\title{Hydrogen-poor Superluminous Supernovae With Late-time H$\alpha$ Emission: Three Events From the Intermediate Palomar Transient Factory}
\author{ Lin Yan$^{1,2}$,  R. Lunnan$^3$, D. A. Perley$^{4,5}$, A. Gal-Yam$^6$, O. Yaron$^6$, R. Roy$^7$, R. Quimby$^{8,9}$, J. Sollerman$^7$, C. Fremling$^7$, G. Leloudas$^{4,6}$, S. B. Cenko$^{10,11}$, P. Vreeswijk$^6$, M. L. Graham$^{12,13}$, D. A. Howell$^{14,15}$, A. De Cia$^{6,16}$, E. O. Ofek$^6$, P. Nugent$^{17, 18}$, S. R. Kulkarni$^{2,3}$, G. Hosseinzadeh$^{14,15}$, F. Masci$^1$, C. McCully$^{14,15}$, U. D. Rebbapragada$^{19}$, P. Wo\'zniak$^{20}$}
\affil{$^1$MS100-22, Caltech/IPAC, California Institute of Technology, Pasadena, CA 91125, lyan@caltech.edu}
\affil{$^2$Caltech Optical Observatories, California Institute of Technology, Pasadena, CA91125}
\affil{$^3$Department of Astronomy, California Institute of Technology, 1200 East California Boulevard, Pasadena, CA 91125}
\affil{$^4$Dark Cosmology Centre, Niels Bohr Institute, University of Copenhagen, Juliane Maries Vej 30, DK-2100, Copenhagen, Denmark}
\affil{$^5$Astrophysics Research Institute, Liverpool John Moores University, IC2, Liverpool Science Park, 146 Brownlow Hill, Liverpool L3 5RF, UK}
\affil{$^6$Benoziyo Center for Astrophysics and the Helen Kimmel Center for Planetary Science, Weizmann Institute of Science, 76100 Rehovot, Israel}
\affil{$^7$The Oskar Klein Centre, Department of Astronomy, Stockholm University, AlbaNova, 10691, Stockholm, Sweden}
\affil{$^8$Department of Astronomy, San Diego State University, San Diego, CA 92182, USA}
\affil{$^9$Kavli IPMU (WPI), UTIAS, The University of Tokyo, Kashiwa, Chiba 277-8583, Japan}
\affil{$^{10}$Astrophysics Science Division, NASA Goddard Space Flight Center, Mail Code 661, Greenbelt, MD 20771, USA}
\affil{$^{11}$Joint Space-Science Institute, University of Maryland, College Park, MD 20742, USA}
\affil{$^{12}$Department of Astronomy, University of Washington, Box 351580, U.W., Seattle, WA 98195-1580}
\affil{$^{13}$ Department of Astronomy, University of California, Berkeley, CA 94720-3411, USA}
\affil{$^{14}$Las Cumbres Observatory, 6740 Cortona Drive, Suite 102, Goleta, CA 93117, USA}
\affil{$^{15}$Department of Physics, University of California, Santa Barbara, Broida Hall, Mail Code 9530, Santa Barbara, CA 93106-9530, USA}
\affil{$^{16}$European Southern Observatory, Karl-Schwarzschild-Strasse 2, 85748 Garching bei M{\"u}nchen, Germany}
\affil{$^{17}$ Lawrence Berkeley National Laboratory, Berkeley, California 94720, USA}
\affil{$^{18}$ Astronomy Department, University of California, Berkeley, 501 Campbell Hall, Berkeley, Ca 94720, USA}
\affil{$^{19}$Jet Propulsion Laboratory, California Institute of Technology, Pasadena, CA 91109, USA}
\affil{$^{20}$Space and Remote Sensing, ISR-2, MS-B244 Los Alamos National Laboratory,
Los Alamos, NM 87545, USA}

\begin{abstract}

We present observations of two new hydrogen-poor superluminous supernovae (SLSN-I), iPTF15esb and iPTF16bad, showing late-time H$\alpha$ emission with line luminosities of $(1-3)\times10^{41}$\,erg\,s$^{-1}$ and velocity widths of (4000-6000)\,km\,s$^{-1}$. Including the previously published iPTF13ehe, this makes up a total of three such events to date. iPTF13ehe is one of the most luminous and the slowest evolving SLSNe-I, whereas the other two are less luminous and fast decliners. We interpret  this as a result of the ejecta running into a neutral H-shell located at a radius of $\sim10^{16}$\,cm.  This implies that violent mass loss must have occurred several decades before the supernova explosion. Such a short time interval suggests that eruptive mass loss could be common shortly before core collapse, and more importantly helium is unlikely to be completely stripped off the progenitor and could be present in the ejecta. It is a mystery why helium features are not detected, even though non-thermal energy sources, capable of ionizing He, may exist as suggested by the O\,II absorption series in the early-time spectra.  Our late-time spectra ($+240$\,d) appear to have intrinsically lower [O\,I]\,6300\,\AA\ luminosities than that of SN2015bn and SN2007bi, possibly an indication of less oxygen ($<$$10\,M_\odot$).  The blue-shifted H$\alpha$ emission relative to the hosts for all three events may be in tension with the binary model proposed for iPTF13ehe. Finally, iPTF15esb has a peculiar light curve (LC) with three peaks separated from one another by $\sim22$\,days. The LC undulation is stronger in bluer bands. One possible explanation is ejecta-circumstellar medium (CSM) interaction.

\end{abstract}


\section{Introduction}

Superluminous supernovae \citep[SLSNe;][]{Gal-Yam2012} are rare stellar explosions, radiating $10 - 100$ times more energy than normal supernovae.  Their extreme peak luminosities and slowly evolving light curves (LC) cannot be explained by standard models based on radioactive decay of $^{56}$Ni.  Although the detailed physics of SLSNe is not understood, a general consensus from published studies is that their progenitors are massive stars, $>30 - 100\,M_\odot$ \citep{Quimby2011,Gal-Yam2009,Smith2007,Ofek2007,Nicholl2014,Yan2015}.  Observations of SLSNe have highlighted our poor understanding of the late stages of massive star evolution, especially mass loss processes. According to standard stellar evolutionary models, massive stars ($>30\,M_\odot$) are thought to have very little hydrogen at the time of supernova explosion \citep{Georgy2012,Smith2014,Langer2012}. However, detections of two types of SLSNe -- one with and one without  H and He (SLSN-II and SLSN-I respectively, Gal-Yam et al. 2012) -- 
illustrate a much more complex picture of massive star evolution, and indicate 
that their massive progenitors must have two distinctly different mass loss histories. Progenitors of SLSN-I lose their H-envelope long before core explosion.  
In contrast, for a progenitor of a SLSN-II, the stripping of its H-envelope must be incomplete, and the bulk of the H-rich medium is still either loosely bound to or very close to the progenitor at the time of the supernova explosion. 

The observational appearance of a SLSN is largely affected by its progenitor mass loss history. Broadly speaking, at the time of explosion, the progenitor star of a SLSN-II still retains a substantial H-envelope, and its early-time spectra show the characteristic H$\alpha$ emission with both narrow and broad components, indicating ejecta interaction with extended, dense H-rich circumstellar medium (CSM), like a SN\,IIn.  In contrast, the progenitor star of a SLSN-I must have lost most of its H and helium material long before the supernova explosion,  and its early-time spectra detect no H and helium features. However, there must be some SLSNe falling between these two simple categories. For example, a progenitor star could retain a small amount of H material and has no substantial CSM. When such a star explodes, its early-time spectrum would have H$\alpha$ emission, but not like SN\,IIn with both narrow and broad components indicating ejecta-CSM interaction. This type of transient may have been detected already, for example, SN\,2008es, SN\,2013hx, PS15br and possibly CSS121015 \citep{Miller2009,Gezari2009, Benetti2014,Inserra2016a}, which show only broad H$\alpha$ emission in the photospheric phase. Another example would be a SLSN-I progenitor which has lost all of the H-envelopes, but only shortly before the supernova explosion. In such a case, the H-rich material would not have enough time to be completely dispersed into the interstellar medium (ISM) and would be located close enough so that when the supernova explodes, the SN ejecta would be able to catch up with this H-shell, and the subsequent interaction would produce broad H$\alpha$ emission in late-time spectra.  Our observation of SLSN-I iPTF13ehe suggests that indeed such events exist \citep{Yan2015}.

Systematic follow-up observations have led to discoveries of  new features from SLSNe-I, including double peak LCs at early times \citep{Nicholl2015, Vreeswijk2016, Smith2016}, and broad H$\alpha$ and [O\,III]\,4363 \&\ 5007\AA\ emission in late-time spectra of SLSNe-I \citep{Yan2015,Lunnan2016}.  
Well sampled light curves of SN\,2015bn and iPTF13dcc have also resulted in discoveries of LC undulations of SLSNe-I, suggesting possible ejecta interaction with H-poor CSM \citep{Nicholl2016,Vreeswijk2016}.

In this paper, we report two new SLSNe-I events, iPTF15esb and iPTF16bad, showing late-time H$\alpha$ emission, similar to iPTF13ehe.  In addition, the LC of iPTF15esb shows strong light curve undulations. This paper reports the new observations and presents a coherent analysis of all three events.  We also discuss the implication for various physical models and the whole SLSN-I population.
Throughout the paper, we adopt a $\Lambda$CDM cosmological model with
$\Omega_{\rm{M}}$\,=\,0.286, $\Omega_{\Lambda}$\,=\,0.714, and $H_0$\,=\,69.6\,$\rm{km}\rm{s}^{-1}\rm{Mpc}^{-1}$ \citep{Planck2015}.

\section{Targets and Observations }
\label{sec_obs}

We discuss a sample with three SLSNe-I discovered by the Intermediate Palomar Transient Factory (iPTF), including two new events (iPTF15esb and iPTF16bad) and one already published event \citep[iPTF13ehe;][]{Yan2015}.  The basic properties and  the coordinates are summarized in Table~\ref{sample}.  These three events are at a similar distance, $z\sim 0.224 - 0.3434$), the median redshift of PTF SLSNe, due to the survey sensitivity limit. 

All three events have the identical Galactic extinction of E($B - V)=$0.04\,magnitude (Schlafly \&\ Finkbeiner 2011). All fluxes are corrected
assuming the extinction law of \citet{Cardelli1989} with $R_V =  A_V/E(B - V) = 3.1$.  The host galaxies have either pre-explosion photometry from SDSS or measurements after the supernova has faded in the case of iPTF13ehe.  The host of iPTF15esb was detected by SDSS and has AB magnitudes of 23.65, 22.61, 21.90, 21.50 and 21.44\,mag in $u,g,r,i,z$ respectively. The host of iPTF16bad was not detected by SDSS in any band, and is fainter than 50\%\ completeness limits  of 22.4, 22.6, 22.6, 21.7, 20.9 in $u,g,r,i,z$ respectively \citep{Abazajian2003}. The absolute $r$ magnitudes are $>-18.5$ and $-18.5$\,mag for iPTF16bad and iPTF15esb respectively.  Compared with $M_r = -21.23$ for a L$^*$ galaxy at $z=0.1$ \citep{Blanton2003}, these two host galaxies are low luminosity dwarfs, typical of SLSN-I host galaxies as found by \citet{Lunnan2014}, \citet{Leloudas2015}, and \citet{Perley2016}. 

Photometric observations of iPTF15esb and iPTF16bad were obtained with the Palomar 48 \&\ 60\,inch (P48 \&\ P60), the 4.3\,meter Discovery Channel Telescope (DCT) and the Las Cumbres Observatory Global Telescope Network (LCOGT).  All reported photometry in Table~\ref{tab:phot} and \ref{tab:phot2} is in AB magnitudes and calibrated to the SDSS $g, r, i$ filters. The P60 and LCOGT photometry is measured using a custom image subtraction software \citep{Fremling2016} and the P48 using the PTF Image Differencing Extraction (PTFIDE) software \citep{Masci2016}.

iPTF15esb and iPTF16bad have spectra at 12 and 4 epochs, covering the rest-frame phase (relative to the peak date) from +0 to +320 and +3 to +242\,days respectively (Table~\ref{tab:spec}).
These data were taken with the Double Beam SPectrograph \citep[DBSP;][]{Oke1982} on the 200\,inch telescope at Palomar Observatory (P200), the Low-Resolution Imaging Spectrometer \citep[LRIS;][]{Oke1995} and the DEep Imaging Multi-Object Spectrograph \citep[DEIMOS;][]{Faber2003} on the Keck telescopes.
The absolute flux calibration of these spectra is set by the broad band photometry at the corresponding phase.

\section{Analysis and Results}
\label{sec_results}

\subsection{Emergence of H$\alpha$ emission from H-poor SLSNe}
\label{sub_halpha}

\begin{figure*}
\plotone{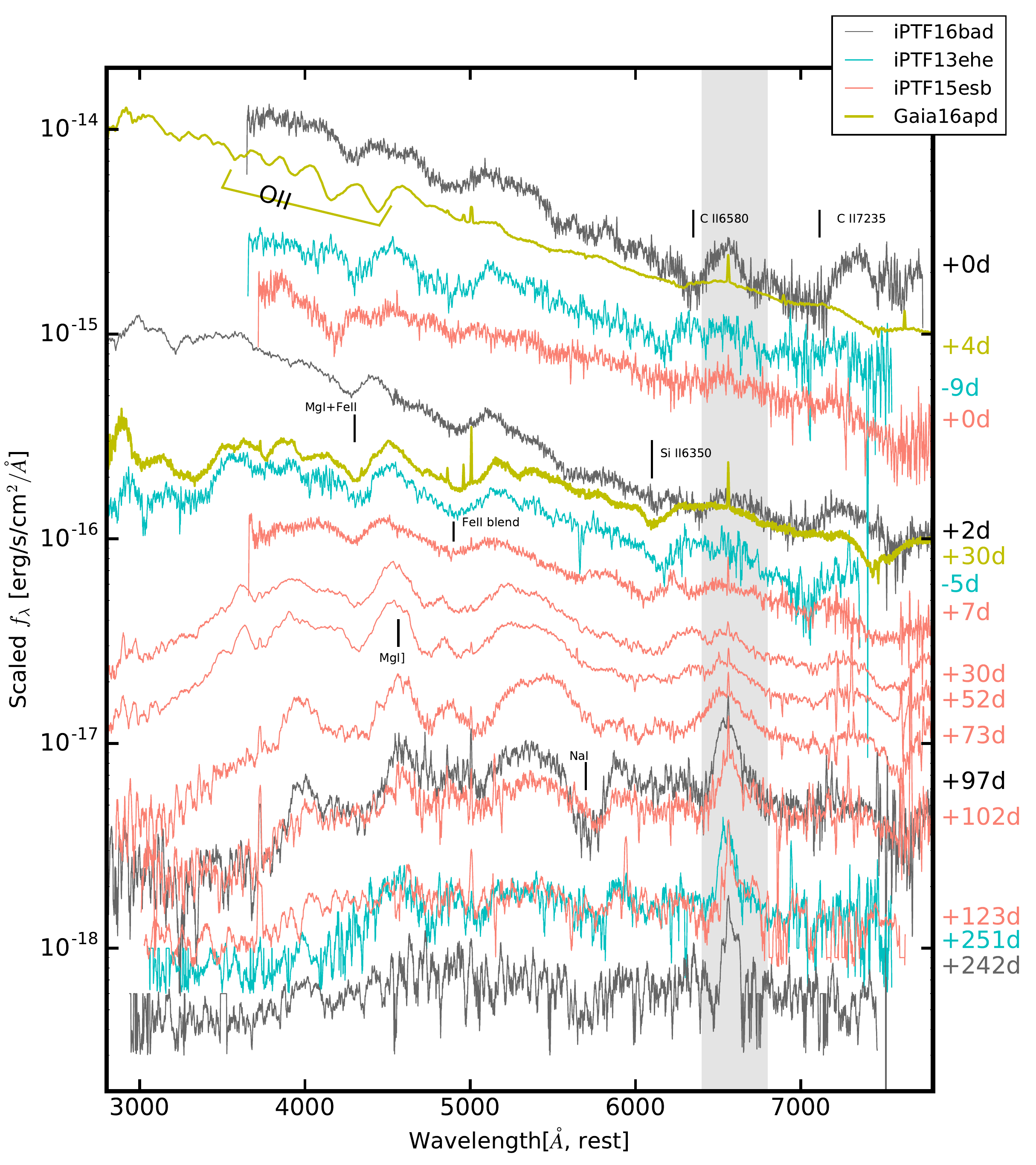}
\caption{The plot displays all of the available spectra from our three SLSNe-I. We also plot the spectrum of SLSN-I Gaia16apd for comparison \citep{Yan2016}.  For visual clarity, we multiplied scaling factors to shift spectra apart, and very noisy pixels near 7800\AA\ are clipped in order to reduce the spectral overlapping.
\label{samplespec}}
\end{figure*}

\begin{figure*}[ht!]
\includegraphics[width=1\textwidth]{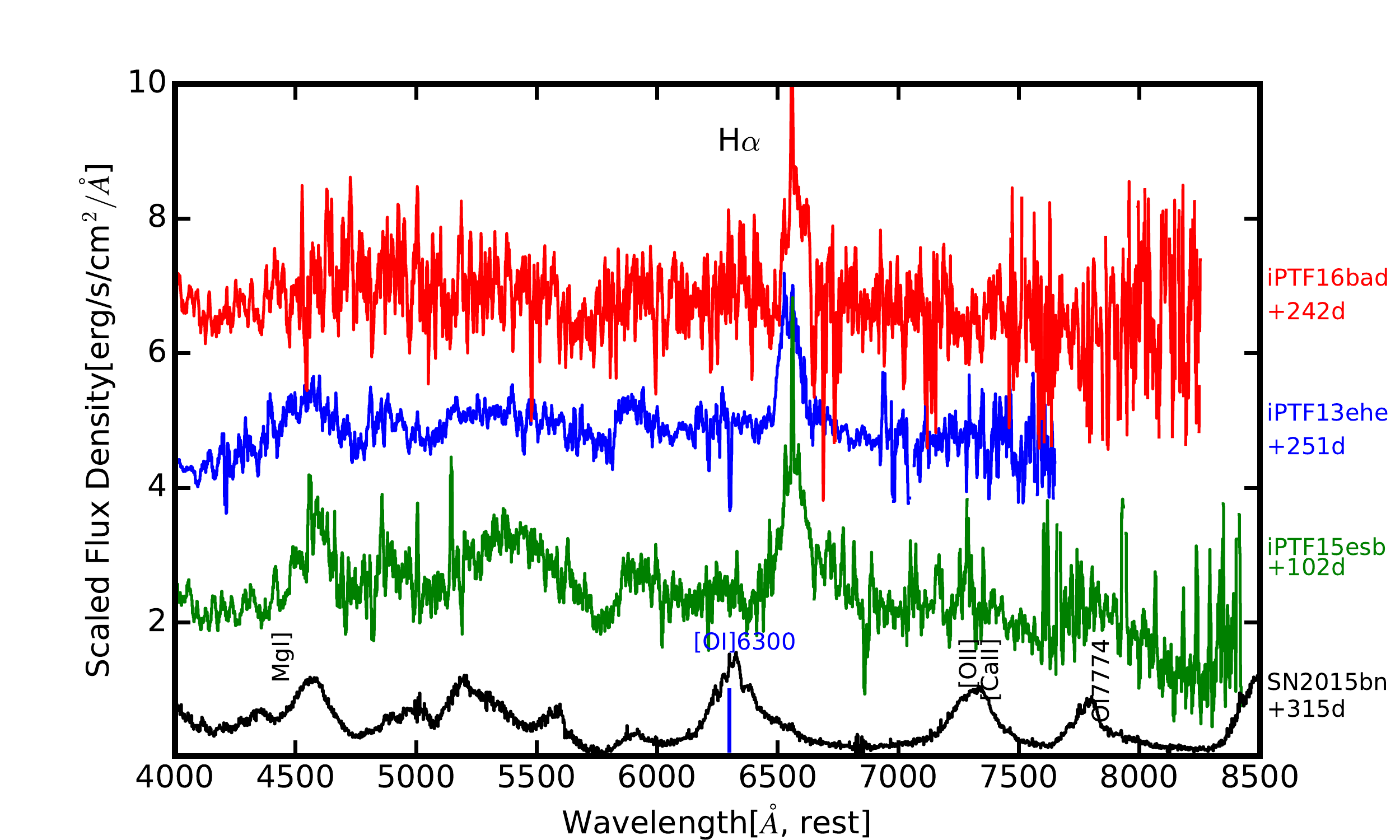}
\caption{The plot displays the late time spectra of iPTF15esb, iPTF16bad, iPTF13ehe, in comparison with that of SLSN-I SN2015bn. Strong H$\alpha$ emission are detected in the spectra of the first three events. 
\label{discoverfig}}
\end{figure*}

The main result of this paper is the detection of broad H$\alpha$ emission in the late-time spectra of the three H-poor SLSNe.
Figure~\ref{samplespec} displays all of the available spectra for these three events, except two spectra of iPTF15esb at +270 and 320\,days, which show only features from the host galaxy.   In this figure, the spectra have not been host subtracted.  It is apparent that broad H$\alpha$ emission lines start to emerge at late-times between photospheric and nebular phases. And they persist until fairly late-times, $+123$, $+242$ and $251$\,days for iPTF15esb, iPTF16bad and iPTF13ehe respectively, as shown in  Figure~\ref{discoverfig}. 
It is worth noting here that the LCs of iPTF15esb and iPTF16bad decline $\sim3$ times faster than that of iPTF13ehe (see \S\ref{subsec_LC} for details). Therefore, the last spectrum from iPTF15esb at $+123$\,d could be at a similar late phase as that of iPTF13ehe.
In addition, Figure~\ref{discoverfig} compares our late-time spectra with the spectrum of SLSN-I SN2015bn \citep{Nicholl2016b}, showing prominent broad H$\alpha$ emission and apparent weak [O\,I]\,6300\,\AA\ lines from our three events.  Quantitative discussion on [O\,I]\,6300\,\AA\ is included in \S\ref{sec_interp}.

One important constraint is when H$\alpha$ is first detected in the available spectra. The answer affects how we calculate the distance the ejecta have traveled since the explosion.  
We display all of the available spectra for these three events in Figure~\ref{samplespec}.  All spectroscopic  data are listed in Table~\ref{tab:spec}, and will be made available via WISeREP \citep{Yaron2012}. For comparison, we also
include the high SNR spectrum of Gaia16apd, which is the second closest SLSN-I ever discovered \citep{Yan2016}. 

It is clear from Figure~\ref{samplespec} that the answer to the above question is not obvious because before a spectrum becomes fully nebular, broad absorption features can make it difficult to determine where the true continuum is.  For examples, does the $+52$\,d spectrum for iPTF15esb (Figure~\ref{samplespec}) have H$\alpha$ emission?  And is the broad bump near 6500\AA\ in the $+0d$\,day spectrum of iPTF16bad H$\alpha$, or continuum between two broad absorption features?  

\begin{figure}[!h]
\includegraphics[width=0.48\textwidth]{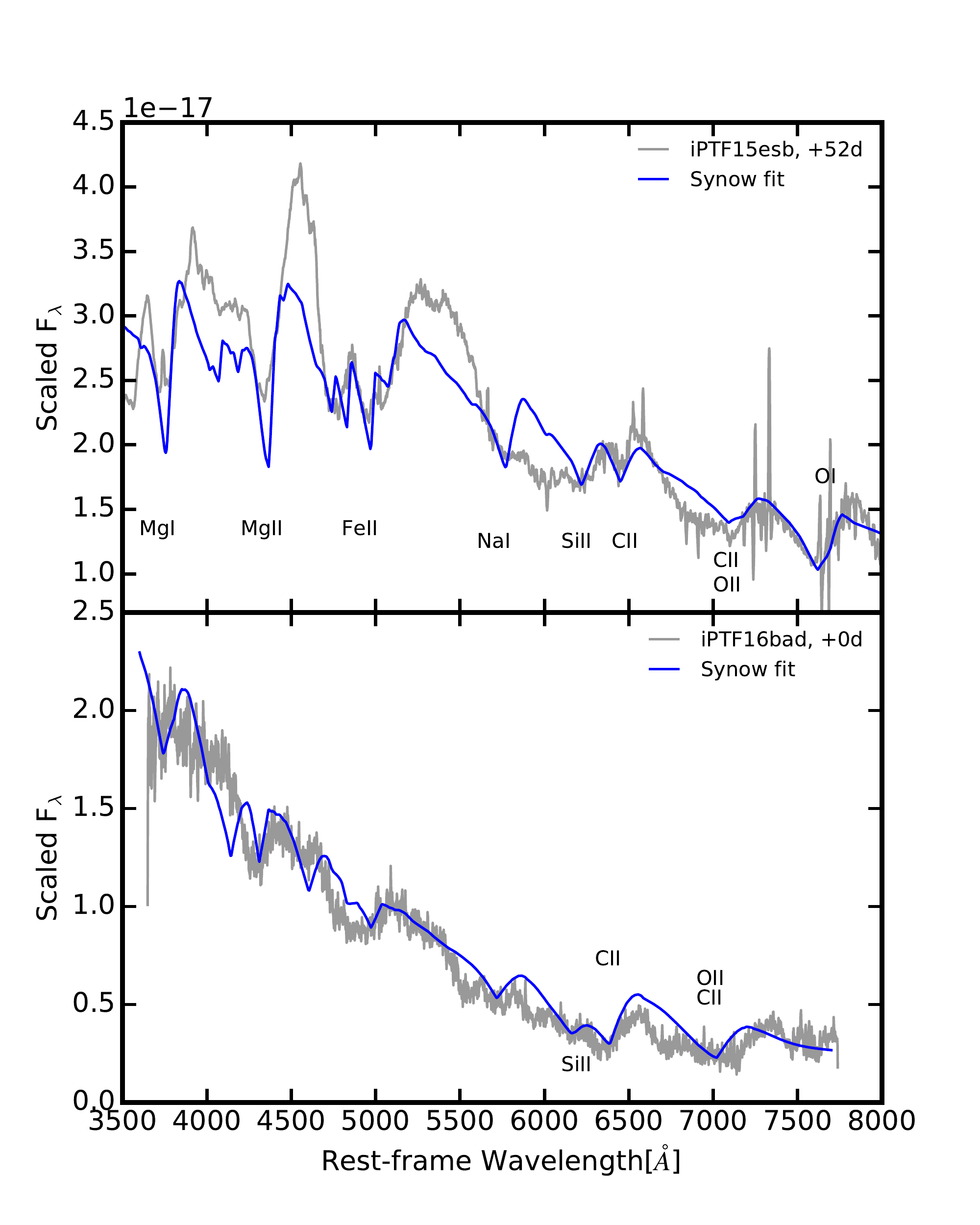}
\caption{Synow model fits to the $+52$\,day and $+0$\,day spectra of iPTF15esb and iPTF16bad. The important wavelength region is near $6300 - 6700$\,\AA\ where the H$\alpha$ emission is.  The apparent features at early times near 6500\,\AA\ are not due to H$\alpha$ emission, but produced by other broad absorptions surrounding that wavelength region. \label{synow}}
\end{figure}

To identify possible absorption features near the 6563\AA\ region, we run SYNOW, the spectral synthesis code \citep{Thomas2013}. This is a highly parametric code, including ion species, temperature, opacity, photospheric velocity, and the velocity distribution. However, it nevertheless provides a useful consistency check for line identifications. Figure~\ref{synow} illustrates the two model fits to the $+52$\,day and $+0$\,day spectra for iPTF15esb and iPTF16bad. Clearly, the observed features near $6000 - 6700$\,\AA\ can be well fit by a combination of Na\,I, FeII\,6299, 6248\AA, Si\,II\,6347, 6371\AA\ and CII\,6580, 7234\AA\ {\it absorption}, without any H$\alpha$ emission. 
This is confirmed by the actual detections of these lines at $+30$\,day in SLSN-I Gaia16apd \citep{Yan2016}.
The broad bumps around 6500\AA\ in post-peak and pre-nebular spectra are also seen in SN\,2007bi and SN\,2015bn \citep{Gal-Yam2009, Nicholl2016}, and are considered to be a result of multiple absorption features.  

By visual inspection of the available spectra, we take $+73$, $+97$ and $+251$\,day as the first dates when H$\alpha$ emission lines are clearly detected.  This method seems to be subjective, however, lack of full spectroscopic coverage gives much larger uncertainties in determining the true times when H$\alpha$ first appears.  

\subsection{Light Curves: Are these three SLSNe-I special?  \label{subsec_LC}} 

\begin{figure}[!ht]
\includegraphics[width=0.48\textwidth,height=0.59\textwidth]{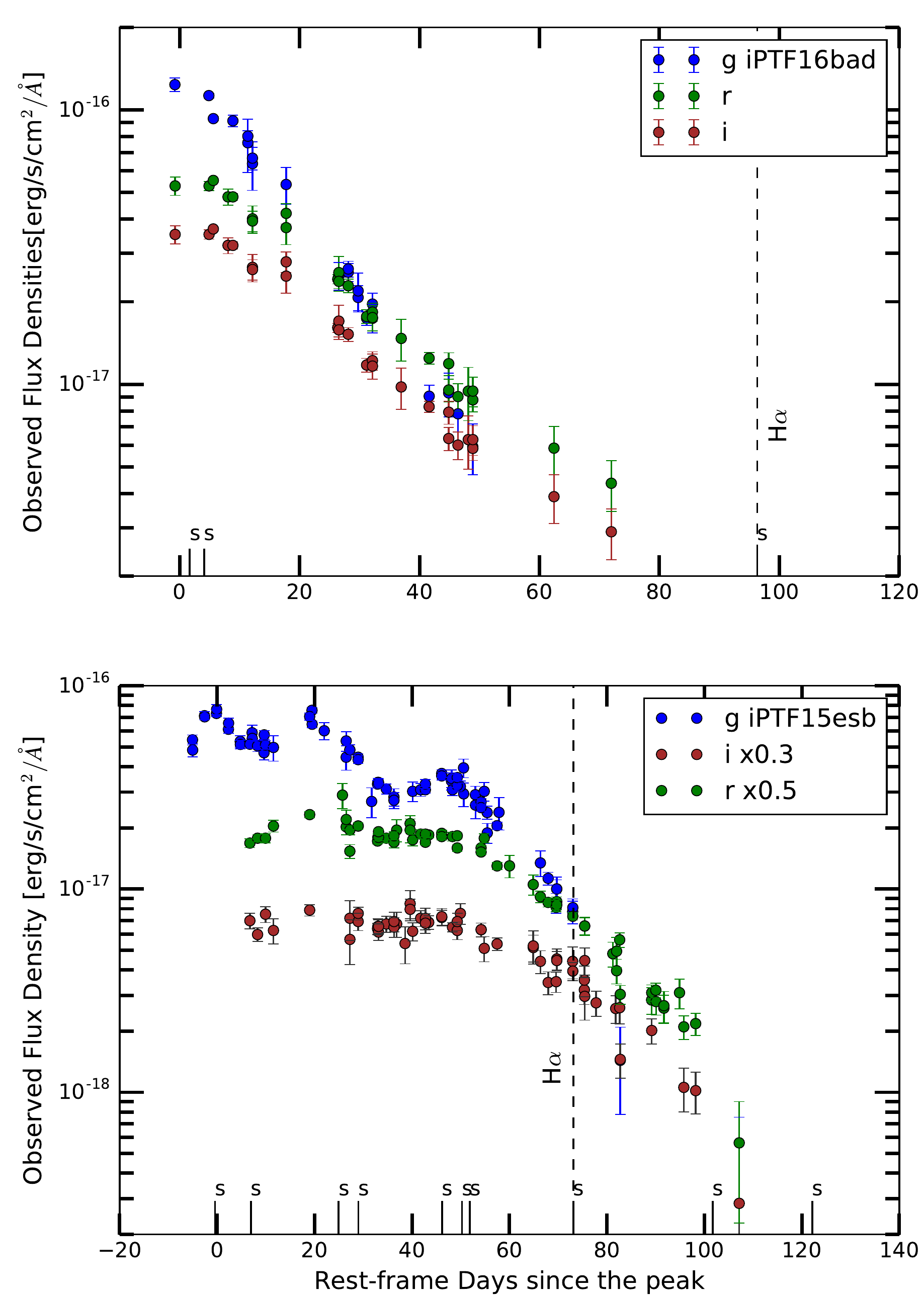}
\caption{This figure presents the monochromatic light curves for $g,r,i$ bands. The Y-axis is observed flux densities in erg/sec/cm$^2/\AA$.  The $r$-band peak date is MJD=57363.5,  57540.4 and 56670.3\,days for iPTF15esb, iPTF16bad and iPTF13ehe respectively.
The black vertical lines at the bottom of the figure mark the dates of spectroscopic observations. The dashed line indicates the date when the H$\alpha$ emission feature starts to appear.  \label{obsLC}}
\end{figure}

\begin{figure}[!h]
\includegraphics[width=0.48\textwidth,height=0.35\textwidth]{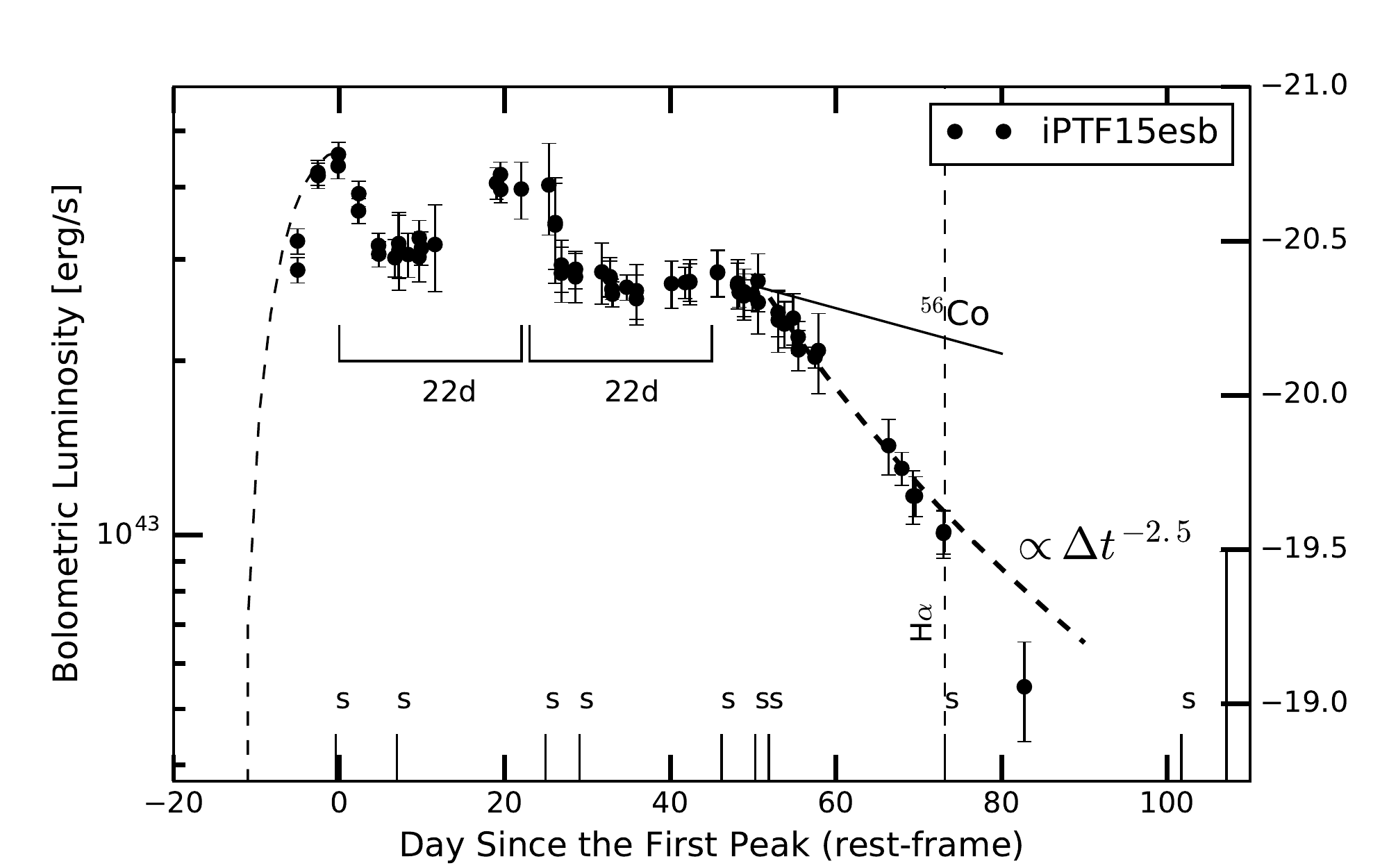}
\caption{The derived bolometric light curve for iPTF15esb. The three peaks are separated from each other by roughly 22\,days.  The very late decay rate is $\propto t^{-2.5}$(dashed line), steeper than the $^{56}$Co rate (solid line). The vertical bars with $s$ at the bottom mark the dates with spectroscopic observations. \label{bolo}}
\end{figure}

\begin{figure}[!h]
\includegraphics[width=0.48\textwidth,height=0.39\textwidth]{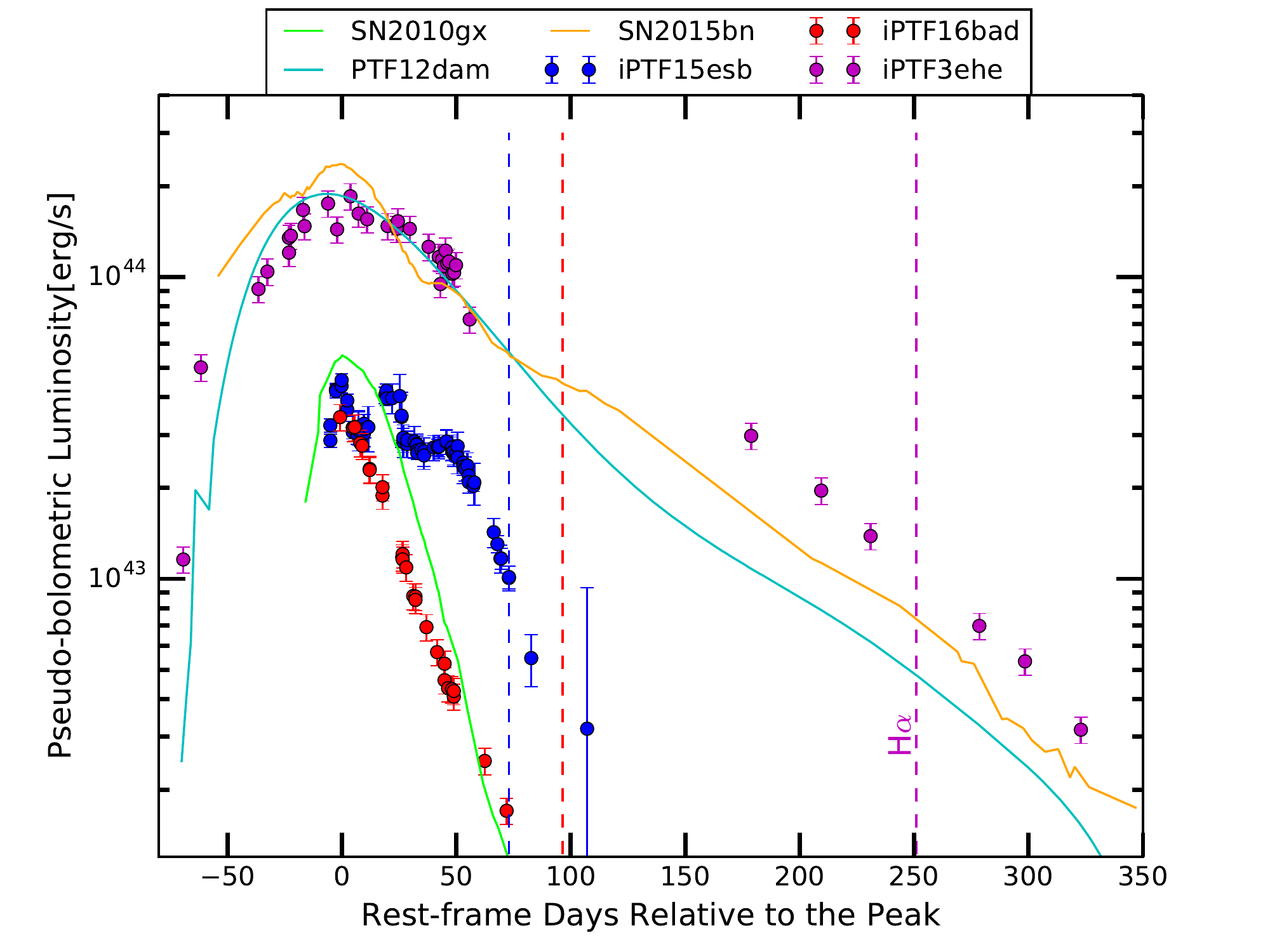}
\caption{ The comparison of the three bolometric light curves.  The dashed vertical lines mark the dates when broad H$\alpha$ emission lines are detected. We also plot the bolometric LCs of SN2010gx (green line), a fast evolving SLSN-I and SN2015bn (yellow) and PTF12dam(cyan), two slow evolving SLSNe-I \citep{Pastorello2010, Vreeswijk2016, Nicholl2016}. \label{3LC}}
\end{figure}

Figure~\ref{obsLC} presents the observed $g,r,i$ light curves of iPTF15esb and iPTF16bad. The derived bolometric light curves are shown in Figure~\ref{bolo} and Figure~\ref{3LC}.  It is immediately clear that the LCs of iPTF15esb are different from a typical SLSN LC, showing prominent undulations, stronger in the bluer bands. The three peaks are roughly separated by $\sim22$\,days. Detailed discussion on the iPTF15esb LC morphology is presented in \S\ref{subsec_15esbLC}.

The peak date for iPTF15esb is chosen as the first peak at MJD\,=\,57363.5\,days. iPTF16bad has very limited photometric data. However, its $r$ and $i$-band LCs in Figure~\ref{obsLC} are initially flat, suggesting we discovered this event just before peak.
Thus we set the peak date as MJD\,=\,57540.4\,days, the epoch of the first data.    
We construct the bolometric light curve for iPTF15esb 
using the following procedure.  We start with a pseudo-bolometric light curve which is an integral of the broad band photometry. At each epoch with a spectrum, we calculate a bolometric luminosity using a blackbody fit. The ratio between the bolometric and pseudo-bolometric luminosity gives the bolometric correction.  Without sufficient early-time photometry, we fit a power-law $L \propto t^2$ form to the pre-peak data points, and derived a minimum $t_{rise} \ge 10$\,days. The late-time decay rate follows $\propto \Delta t^{-2.5}$, much steeper than the $^{56}$Co decay rate (solid line in Figure~\ref{bolo}). 
The bolometric LC for iPTF15esb is shown in Figure~\ref{bolo}.  The similar method was used for iPTF13ehe to get the bolometric light curve. iPTF16bad does not have many spectra. We derive its bolometric LC by assuming similar bolometric corrections to the pseudo-bolometric LC as that of iPTF15esb.

One important question is: are these three SLSNe-I with late-time H$\alpha$ emission special and have distinctly different photometric properties compared to other SLSNe-I? The answer is relevant to understanding the nature of these events.  Figure~\ref{3LC} makes comparison of these three LCs with other events, including two slow evolving SLSN-I PTF12dam, SN2015bn and one fast evolving SLSN-I SN2010gx \citep{Vreeswijk2016, Nicholl2016, Pastorello2010}.

iPTF15esb and iPTF16bad have peak bolometric luminosities of $\sim$4$\times10^{43}$\,erg\,s$^{-1}$ ($-20.57$\,mag), whereas iPTF13ehe is more energetic, with $L_{peak}$$\sim$$1.3\times10^{44}$\,erg\,s$^{-1}$ ($-21.6$\,mag\footnote{Here we assume a solar bolometric magnitude of $4.74$.}).
Although an unbiased SLSN-I sample does not yet exist,  a simple compilation of 19 published SLSNe-I \citep{Nicholl2015} has a median $<L_{peak}> \sim 5.7\times10^{43}$\,erg\,s$^{-1}$ ($-20.7$\,mag).  
  
In addition, one striking feature in Figure~\ref{3LC} is the large difference in evolution rates between the three LCs.  For iPTF15esb and iPTF16bad, their post-peak decay rates are fast, $\sim 0.05$\,mag/day, 3 times faster than that of iPTF13ehe, which is $0.016$\,mag/day. For comparison, $^{56}$Co decay rate is 0.0098\,mag/day. 
The LC evolution of iPTF15esb and iPTF16bad is similar to the fast evolving SLSN-I SN2010gx \citep{Pastorello2010}, and iPTF13ehe is more like the extremely luminous, slowly evolving SLSN-I SN2007bi \citep{Yan2015,Gal-Yam2009}.  For a naive comparison with the compiled SLSN-I sample by \citet{Nicholl2015},  67\%\ have decay rates of $0.03 - 0.05$\,mag/day, and 33\%\ of $0.01-0.02$\,mag/day.  In addition, iPTF13ehe has a rise time scale of $83-148$\,days, implying a large ejecta mass of $70-220\,M_\odot$. The other two events do not have sufficient pre-peak data, but $t_{rise}$ in iPTF15esb is likely short, as suggested by the rising rate of the first two available observations before the peak.

We conclude that the photometric properties of these three events are clearly very different from each other. However, they are within the diverse ranges represented by the published SLSNe-I so far, with the possible exception of the unique LC morphology  of iPTF15esb. Therefore, it is possible that whatever physical processes responsible for late-time H$\alpha$ emission could also be relevant to the whole population.

\subsection{Other Spectral Properties}
In the following sections, we describe other properties measured from the full spectral dataset.
\subsubsection{Rapid Spectral Evolution}

As shown in Figure~\ref{samplespec}, the spectra of these three events are similar to each other at both early and late-time.  However, they are very different from the spectra of Gaia16apd in two aspects.  First, our spectra at maximum light do not have the full O\,II absorption series (5 features, as seen in Gaia16apd) at 4000\AA, the hallmark of a typical SLSN-I at early phases \citep{Quimby2011}.  Instead, their absorption features at 4200\AA\ could be one or two features of the full O\,II absorption series. This is supported by the matching between the early-time spectrum of Gaia16apd and that of our events. However, we caution
that in iPTF16bad, this identification of the partial O\,II absorption is uncertain. More detailed analyses and modelings are discussed in \citep{Quimby2017} and \citep{Gal-Yam2017}.

As pointed out by \citet{Mazzali2016},  O\,II absorptions arise from highly excited O$^+$ with an excitation potential of $\sim25$\,eV.  Such a high energy level implies that the excitation of O\,II levels is not in thermal equilibrium with the local radiation field. For example, $\gamma$-ray photons from radioactive decays could be a source of excitation energy. This effect is generally represented by a tunable parameter in models, for more details see \citet{Mazzali2016}.

The second prominent difference is that spectroscopically, iPTF13ehe and iPTF15esb seem to evolve faster than Gaia16apd, developing strong Mg\,I and Fe\,II blends at $-5$ and $+7$\,days, characteristics of a SLSN-I at later times, such as the Gaia16apd spectrum at $+30$\,day.  Naively, this may seem to suggest a lower ejecta mass because less material could cool down faster. This may be the case for iPTF15esb, but is not correct for iPTF13ehe at all because the slow rise time of its LC requires a very high ejecta mass \citep{Yan2015}.  This suggests that spectral evolution is affected by many other factors.
The situation for iPTF16bad is not clear due to lack of sufficient spectroscopic data.  

\subsubsection{Higher Ejecta Velocities}
Figure~\ref{veltime} shows the ejecta velocity and the blackbody temperature as a function of time.
When possible, we use FeII\,5169\AA, a commonly used feature, to measure the velocity evolution with time. Other Fe\,II lines, such as FeII\,4924, 5018, 5276\AA\ are also used to cross check the results, as what is done in \citep{Liu2016}. 
The exception is iPTF15esb at +0\,day, whose spectrum does not have a strong FeII absorption, and the ejecta velocity is estimated using O\,II. The O\,II feature in iPTF15esb is blue-shifted by $40$\AA\ relative to that of PTF09cnd at $-30$\,day with a velocity of 15000\,km\,s$^{-1}$ \citep{Quimby2011}. This implies the velocity of iPTF15esb at $+0$\,day is roughly 17,800\,km\,s$^{-1}$.  The same method is applied to iPTF13ehe and iPTF16bad.  We find that at maximum light, our three SLSNe-I have higher ejecta velocities than those of other published SLSNe-I, ranging between $9000 - 12000$\,km\,s$^{-1}$ \citep{Nicholl2015}.

\begin{figure}[h]
\includegraphics[width=0.47\textwidth,height=0.4\textwidth]{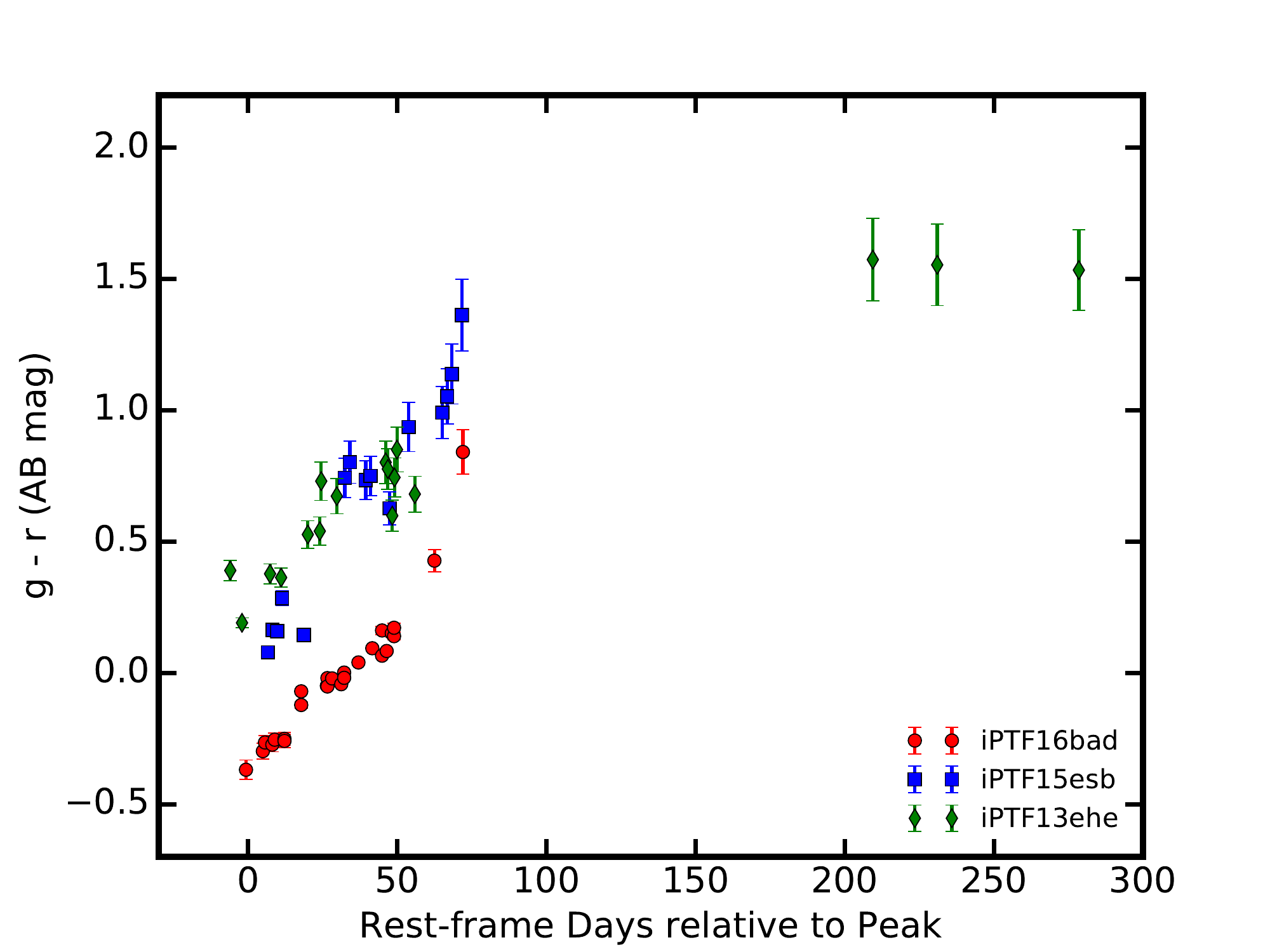}
\caption{The broad-band $(g-r)$ color as a function of time, showing the time evolution of the spectral slopes. \label{colors}}
\end{figure}

\begin{figure}
\includegraphics[width=0.5\textwidth,height=0.55\textwidth]{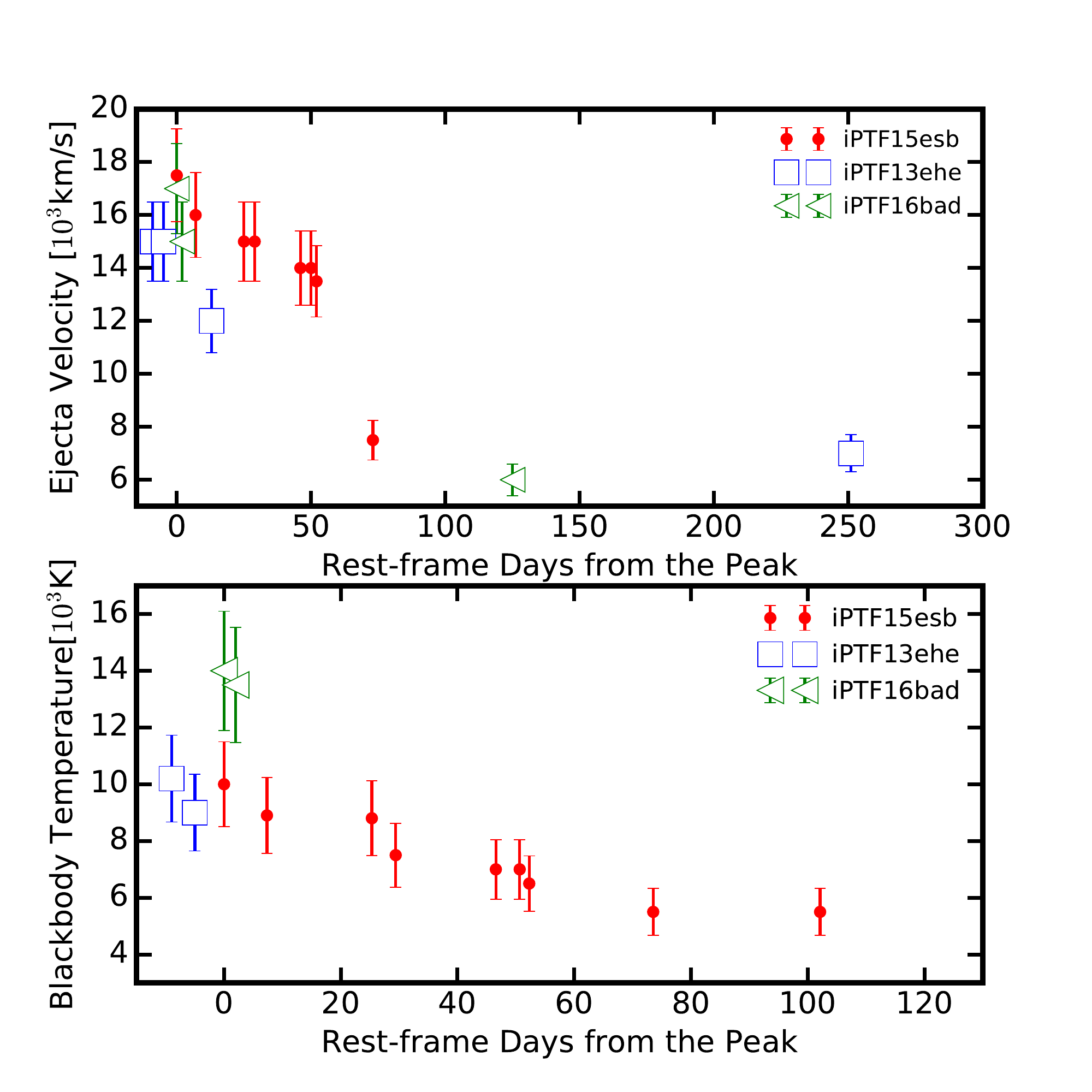}
\caption{Blackbody temperature and ejecta velocity as a function of time. The velocity is measured from FeII\,5169\AA. The details are discussed in the text.
\label{veltime}}
\end{figure}

The blackbody temperatures ($T_{BB}$) are estimated by fitting a blackbody function to the spectral continua.  
At maximum light, the blackbody temperatures of these three SLSNe-I range from $\sim8000 - 14000$\,K, with iPTF13ehe being the coolest whereas iPTF16bad the hottest.  Compared with other SLSNe-I with strong O\,II absorption series such as PTF09cnd, SN2015bn, PTF11rks and Gaia16apd with maximum light $T_{BB} \sim 13000 - 15000$\,K \citep{Inserra2013,Quimby2011,Nicholl2016,Yan2015},  iPTF15esb and iPTF13ehe are indeed cooler at peak phase.  At the peak phase, iPTF16bad has a hotter temperature, which is shown by its steeper and bluer spectra in the early-times (Figure~\ref{samplespec}). This difference is confirmed by their broad band $(g - r)$ color versus time shown by Figure~\ref{colors}.  
The lack of the full O\,II absorption series should not be associated with blackbody temperatures at the peak phase.  This is because the excitation of O\,II levels are certainly non-thermal. 

%%%%%%%%%
\subsubsection{Broad H$\alpha$ Line Luminosities and  Velocity Offsets}

We perform simultaneous spectral fitting to the spectral continuum plus both narrow and broad H$\alpha$ components assuming a Gaussian profile. The narrow line fluxes are iteratively measured from the unsmoothed data.
Table~\ref{tab_linefit} lists the measured line luminosities and velocity widths.

\begin{figure}
\includegraphics[width=0.5\textwidth,height=0.6\textwidth]{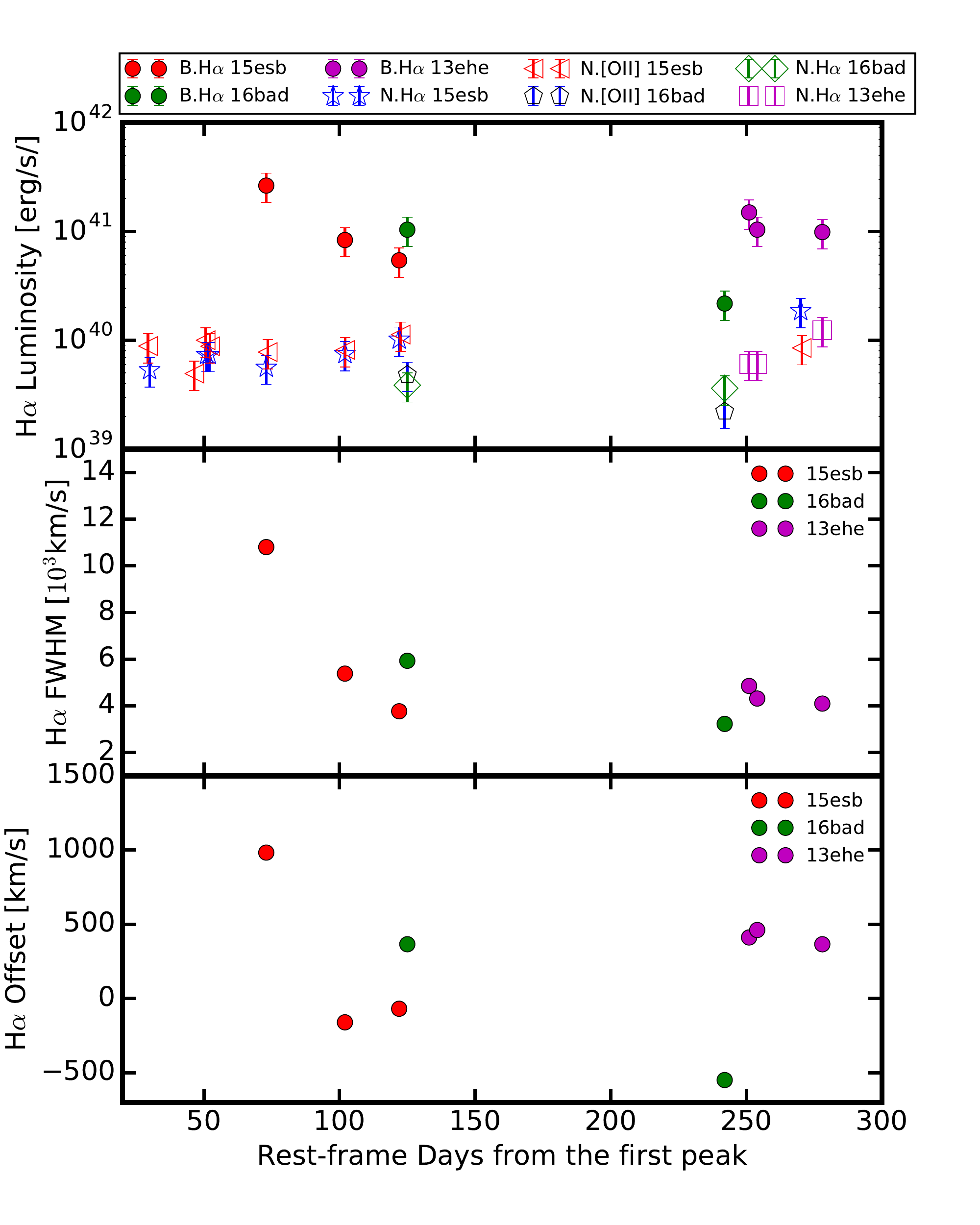}
\caption{In three panels, we plot the H$\alpha$ line luminosities, line widths (FWHM) and line centroid shifts in km\,s$^{-1}$ as a function of time for the three events discussed in this paper. The X-axis shows the time in days in the rest-frame, relative to the peak date for each SLSN-I.  
\label{Halpha-flux}}
\end{figure}

One important question is whether the narrow H$\alpha$ line comes from the host or from the supernova.  The top panel in Figure~\ref{Halpha-flux} shows the integrated line luminosities as a function of time.  The narrow H$\alpha$ and [O\,II] lines show slight variations with time, and the changes are less than a factor of 2. 
In contrast, the broad H$\alpha$ line luminosities vary by a factor of 10 with time. Furthermore, the centroids of the narrow H$\alpha$ emission are always at 6563\AA, whereas the centroids of the broad components change with time (see below).  In the case of iPTF13ehe, the spatially resolved 2-D spectrum shows that the narrow emission appears to be at the center of the host galaxy \citep{Yan2015}. We conclude that the narrow H$\alpha$ and [O\,II] emission lines are likely dominated by the host galaxies. Narrow H$\alpha$ emission from the supernovae may exist, but is too low luminosity to be detected by our data. The observed small variations are due to the combined effects of variable seeing and slit losses. 
 
The middle panel in Figure~\ref{Halpha-flux} shows the broad H$\alpha$ line width (FWHM) as a function of time.  The FWHM of $\sim 6000 - 4000$\,km\,s$^{-1}$ should not be interpreted as the shell expanding velocity.  Similar to well studied SNe\,II powered by ejecta interaction with Circumstellar Medium (CSM),  the broad line widths likely indicate the velocities of the shocked material.  The H-rich CSM expansion velocity is probably much smaller, of an order of a few 100\,km\,s$^{-1}$.    

These three events show an interesting trend in their velocity offsets between the broad and narrow H$\alpha$ components, as shown in the bottom panel in Figure~\ref{Halpha-flux}.  We find that initially the broad components appear to be blue-shifted relative to the narrow components (assuming host emission), and at later-times, become red-shifted.  The velocity offset for iPTF15esb at $+73$\,days is as high as +1000\,km\,s$^{-1}$, and decreases to $\sim -400$\,km\,s$^{-1}$ at later epochs.  Similarly in iPTF16bad, the offset varies from $+400$ to $-500$\,km\,s$^{-1}$ at +125 and $242$\,days. iPTF13ehe shows only positive velocity offsets (blue-shifted). 

\begin{figure}
\includegraphics[width=0.48\textwidth]{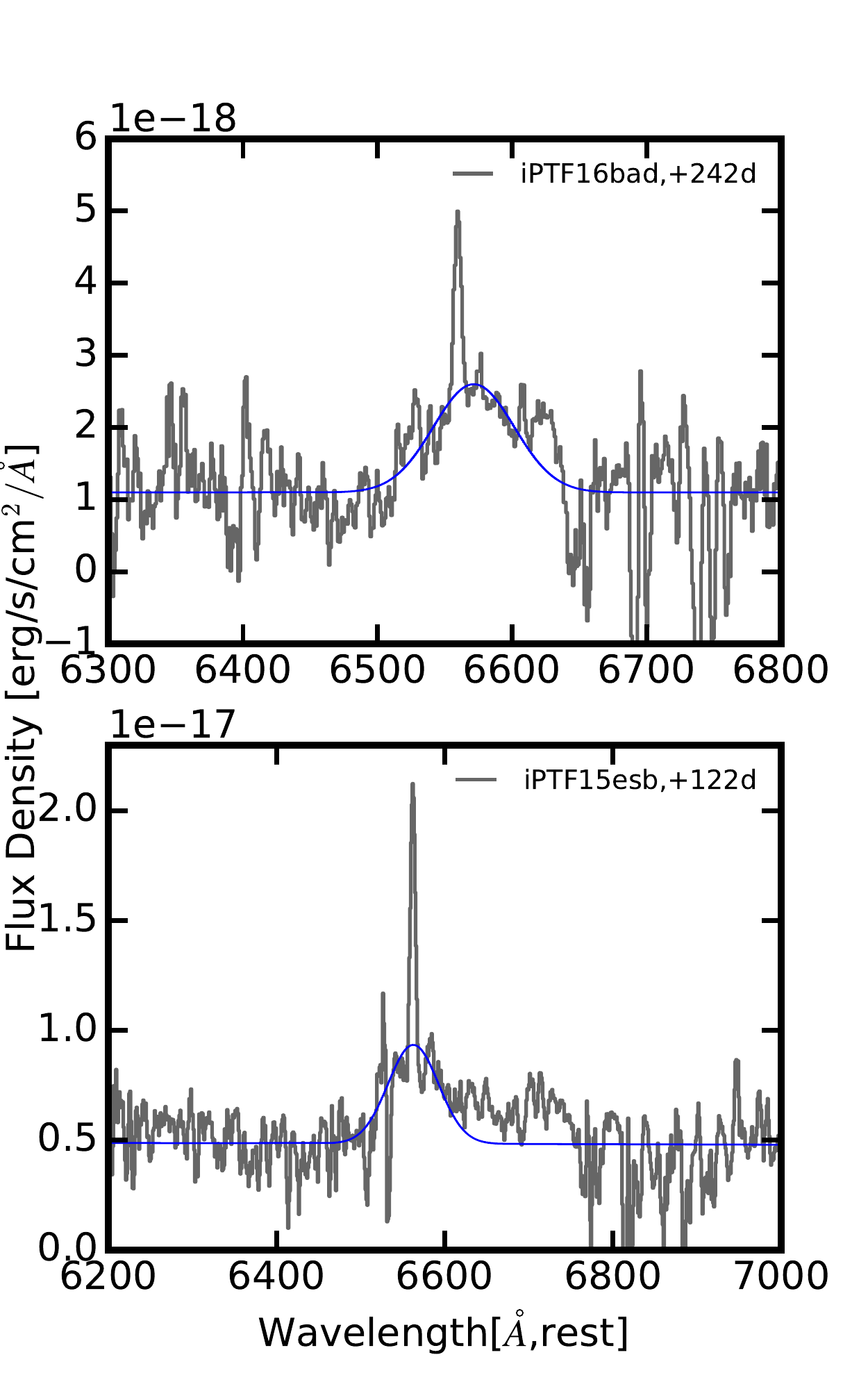}
\caption{ Two spectra for iPTF15esb and iPTF16bad showing some excess emission red-ward of H$\alpha$ 6563\,\AA.  \label{redwing}}
\end{figure}

Figure~\ref{redwing} shows the +122\,d and +242\,d spectra for iPTF15esb and iPTF16bad.  Although noisy, the spectra show the excess emission at the red side of H$\alpha$ 6563\,\AA.  One possible explanation  is that initially the expanding H-shell could obscure the H$\alpha$ photons from the back side which is moving away from us. So we initially see more H$\alpha$ emission from the material moving toward us (blue-shifts).  In this model, at later-times when ejecta become more transparent, we should see more symmetric line profiles with no velocity offsets.  This is clearly not what we see at very late-times in our data.  So the red excess emission can not be explained by obscuration.
We also note that the observed positive velocity offsets in all three events could be in tension with the binary model proposed by \citet{Moriya2015} for iPTF13ehe, which predicts the equal probability of observing both positive and negative velocity offsets relative to the host galaxies.

\section{Implications and Discussions }
\label{sec_interp}

\subsection{Nature of the H$\alpha$ emission and implication for helium in the ejecta}

One important question is where this hydrogen material producing the late-time H$\alpha$ emission is located.  One possibility is a residual hydrogen layer left over from incomplete stripping of the H-envelope. 
This scenario can be ruled out because if there is any hydrogen in the ejecta, it is very difficult not to have any H$\alpha$ absorption at all near maximum light, as shown in the detailed modelings carried out by \citet{Hachinger2012}.  This is because ejecta density is usually much higher than $10^{7}$\,cm$^{-3}$, and the H recombination time scale is $\propto 10^{13}/n_e$\,seconds $\sim 11(n_e/10^{7})^{-1}$\,days, quite short. 
An actual example is SN\,1993J, which is a type IIb SN with a very low mass of H ($<0.9M_\odot$). Although its pre-peak spectra have high blackbody temperatures, a weak H$\alpha$ emission is present at early phases \citep{Filippenko1993}. 
For our three events, at post-peak before $+70$\,days, the photosphere temperatures are already low and there are no detectable H$\alpha$ features in all three events, even with ample spectral coverage between $+0$\,days and $+122$\,days for iPTF15esb.

The second possibility is a neutral, detached H-rich shell located at a distance from the progenitor star, possibly produced by a violent mass loss episode some time prior to the supernova explosion.  When the ejecta eventually run into this shell, the shock interaction ionizes H atoms, and subsequent recombination produces H$\alpha$ emission.  This idea was initially proposed for iPTF13ehe in \citet{Yan2015}.  

The third possibility is proposed by \citet{Moriya2015}, where the progenitor is in a binary system with a massive, H-rich companion star ($>20M_\odot$). The explosive ejecta strip off a small amount of H from the companion star, which is then mixed in the inner regions of  the ejecta. This H-rich material becomes visible only when the inner layers of ejecta are transparent in the nebular phases.  This model predicts that depending on the orientation of the binary, there should be equal probability of seeing H-emitting material moving toward or away from us. The fact that we see blueshifts in all three events could be in tension with this prediction and disfavors this model. 

Some quantitative parameters for the H-shell model can be derived from our observations. 
In this scenario, the progenitor is a massive star, prone to violent mass losses.
Several decades before the explosion, it undergoes an eruptive episode,  ejecting all of the remaining H envelope.  Let us assume that the ejecta average speed is $<$$v_{ej}$$>$ and the time between the explosion and the time H$\alpha$ is first detected is $\Delta t$. The distance traveled by the ejecta, {\it i.e.} the radius of this shell, is thus $R = 8.6$$\times$$10^{15}({v \over 10^4\,km/s})$$\times$$({\Delta t \over 100d})$\,cm.   Our measured ejecta velocities range from $18000 - 15000$\,km\,s$^{-1}$ at maximum light to $\sim 6000 - 7000$\,km\,s$^{-1}$ at $+100$ to $+250$\,days.  So the baseline assumption of $<$$v_{ej}$$>\sim10,000$\,km\,s$^{-1}$ is not too far off. 

The values of $\Delta t$ are not well measured for both iPTF15esb and iPTF16bad due to poorly constrained explosion dates. For iPTF15esb, a rough estimate of $\Delta t$ (Figure~\ref{obsLC}) is $73+20 \sim 93$\,days.  iPTF13ehe is a slowly evolving SLSN-I \citep{Yan2015}, with $\Delta t \sim 332$\,days.  Therefore, for these three events, the sizes of the H-rich shells range between $9 - 40\times 10^{15}$\,cm.  If the shell expansion speed is 100\,km\,s$^{-1}$, the time since the last episode of mass loss before explosion is $t_{erupt} \sim R/v_{shell} \sim R/100$\,km\,s$^{-1} \sim 30$\,yrs.  So approximately, the last episode of mass loss is only 30 years before the supernova explosion. This time could be as short as 10 years if the expansion speed is faster.  Such violent instabilities shortly before the supernova explosion could be a very common phenomenon for massive stars in general, as demonstrated for example by flash-spectroscopy of SN\,IIP iPTF13dqy and precursor outbursts in SNe\,IIn such as SN2009ip \citep{Yaron2017, Ofek2014, Margutti2014,Martin2013}.  At lower luminosities,  a similar and well studied example as our three events is SN2014C, which was initially discovered as an ordinary SN Ib, then evolved into a strongly interacting SN IIn over $\sim1$ year time scale \citep{Milisavljevic2015, Margutti2017a}. 

Two commonly asked questions are:  (1) if this H-shell is present, why don't we see any H emission in the early-time spectra?  (2) why are our spectra not being obscured or absorbed by this shell?
This H-shell is likely to be neutral but optically thin during early-times. However, 
if spectra were taken just hours after explosion, neutral H should have been ionized and we would have detected flash spectral features \citep{Gal-Yam2014,Yaron2017}, including H$\alpha$ emission .  For these three events, we don't  detect any early-time H$\alpha$ emission because by the time that our first optical spectrum is taken, the H-shell has already recombined. At a density of $n\geq10^7$\,cm$^{-3}$, the H recombination time $t_{rec} \leq 10^{13}/n$\,sec $\leq11$\,days.  For a H-shell with $R \sim 10^{16}$\,cm and a width of 10\%\,R, this density limit corresponds to a mass limit of $\geq0.01M_\odot$.

If the H-shell is neutral at pre-peak,  why don't we observe any H$\alpha$ absorption? H$\alpha$ absorption is produced when an excited H atom at $n=2$ absorbs a photon with $\lambda=6563$\AA, and moves up to $n=3$ level.  At temperatures of several thousands degree, most H atoms are in the ground state ($n=1$) because the excitation energy from $n=1$ to $n=2$ requires 10\,eV, implying a much higher temperature (100,000\,K). Without excited $n=2$ H atoms, there is no H$\alpha$ absorption. However, we predict that Ly$\alpha$ absorption ($n=1$ -$>$ $n=2$ transition) should be strong. Future late-time UV spectroscopy may confirm this for events such as ours.

The mass of this H-shell can be constrained by two other factors. When the ejecta run into the shell, H atoms are ionized again by the thermalized kinetic energy. One constraint is that this ionized H-rich CSM can not have very high electron scattering opacity, {\it i.e.} Thomson scattering opacity,  $\tau_{thomson} = \sigma_T n_e \Delta R \leq 1$ with $\Delta R$ being the width of the shell. Otherwise,  photons from the central supernova would have been absorbed.  This condition implies $M_{shell} \leq {4\pi f m_H R^2 \over \sigma_T}$, and $f$ is the filling factor, $R$ is the radius of this shell. Here $\Delta R$ is cancelled out when computing the total mass.
With the Thomson cross section $\sigma_T = 6.65\times10^{-25}$\,cm$^{2}$, we have $M_{shell} \leq 1.6f({R\over10^{16}\,cm})M_\odot$.  Assuming the width of this shell is only 10\%\ of the radius $R$, the implied electron volume density $n_e \sim 7\times10^9 ({R\over1.0^{16}\,cm})$\,cm$^{-3}$.  The H-shell upper mass limits range from $(1.6 - 30)f M_\odot$ for the three events discussed in this paper. In the case of a small filling factor $f\sim10\%$, the shell mass would be less than $0.2-3M_\odot$.  One scenario which could naturally explain such a powerful mass loss is the Pulsational Pair-Instability model \citep[PPISN;][]{Woosley2007, Woosley2016}.  Further support of this model is from the weak [O\,I]6300\AA\ emission in the nebular phase spectra, as discussed below.

In the H-shell scenario, the time interval between the supernova explosion and the mass loss episode which ejected all of the H-envelope is not very long, only several decades. During this period of time, additional mass loss episodes could remove some helium layers from the progenitor star, but it is very unlikely that all of the helium can be completely removed.  For example, \citet{Woosley2002} presented a model for a star with $M_{ZAM} \sim 25M_\odot$. Their Figure 9 shows before the supernova explosion, the most outer layer has roughly $5M_\odot$ of mixture of H and He, and just underneath that, there is a pure He layer with a mass of $\sim 1.5M_\odot$.   If we assume the mass loss rate of $10^{-4}$ to $10^{-6}M_\odot$/yr, similar to nominal wind mass loss rates, the time required to completely remove the pure He layer is $1.5\times(10^4  - 10^6)$\,yrs, much longer than several decades set by our observational constraint.

\begin{figure}
\center{
\includegraphics[width=0.49\textwidth]{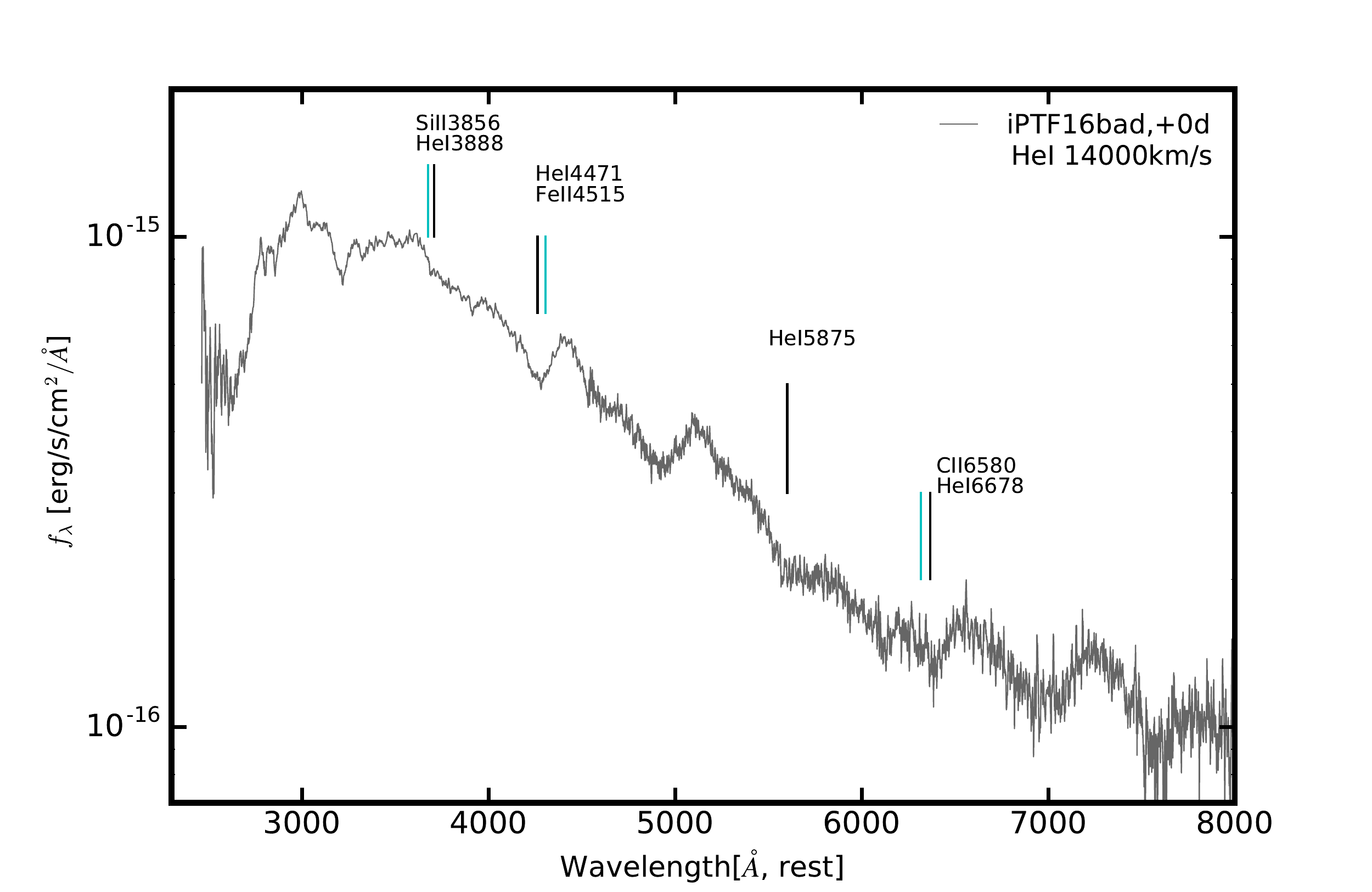}
}
\caption{ The iPTF16bad spectrum at $+0$\,day. We show that weak helium features are difficult to  rule out because they could be blended with other stronger features. As shown, HeI and other features are blued shifted by 14000\,km\,s$^{-1}$. \label{16badHe}}
\end{figure}

This implies that the ejecta of our three events may contain helium. Observationally, our early-time spectra do not detect significant helium features. However,  we caution that presence of weak helium absorption features is very difficult to confidently rule out because He\,I3888, 4417 and 6678\AA\ lines tend to be blended with other features such as strong Fe\,II4515\AA\, Si\,II3856, and C\,II6580\AA, as shown in Figure~\ref{16badHe}.  On another hand, the non-detection of helium features might not be very surprising because helium ionization potential is high, 24.58\,eV. This would require much higher temperatures than what our spectra show, or more likely,  non-thermal ionization conditions, for example, mixing with radioactive material such as $^{56}$Ni.   

Indeed, this condition for non-thermal ionization of He probably exist for SLSNe-I, as suggested by commonly detected five O\,II absorption series around 4000\,\AA. 
As argued in \citet{Mazzali2016},  the excitation of O\,II levels is from non-thermal process, such as  energetic particles from radioactive decays. What is relevant here is that these particles can also ionized He\,I (ionization potential of 24.6\,eV). So if helium is present in the ejecta, it is a puzzle why we do not detect any spectral signatures in the early-time data.

Another possible explanation for weak or absence of He features is that all helium is mixed into the outer H-envelope and the H+He outer layers were completely stripped off the progenitor stars before the supernova explosions. The ejecta contain no helium material.

\subsection{Weak [O\,I]\,6300 doublet emission}

The [O\,I]\,6300 doublet emission is usually very strong for SLSNe-I and SNe\,Ic because of the following two reasons. First, these supernovae are thought to result from the explosions of massive C+O cores.  The ejecta should naturally contain a lot of oxygen.
 Second, the [O\,I]\,6300 line is a very efficient coolant in the nebular phase \citep{Jerkstrand2017}.  Therefore, it is common to see very strong [O\,I]\,6300\,\AA\ emission in core collapse SNe.
Figure~\ref{discoverfig} visually illustrates an apparent lack of [O\,I]\,6300\,\AA\ emission in the late time spectra of our three events. Given the noise level in our spectra, it is not immediately clear whether this apparent discrepancy is significant, however.

\begin{figure}
\center{
\includegraphics[width=0.49\textwidth]{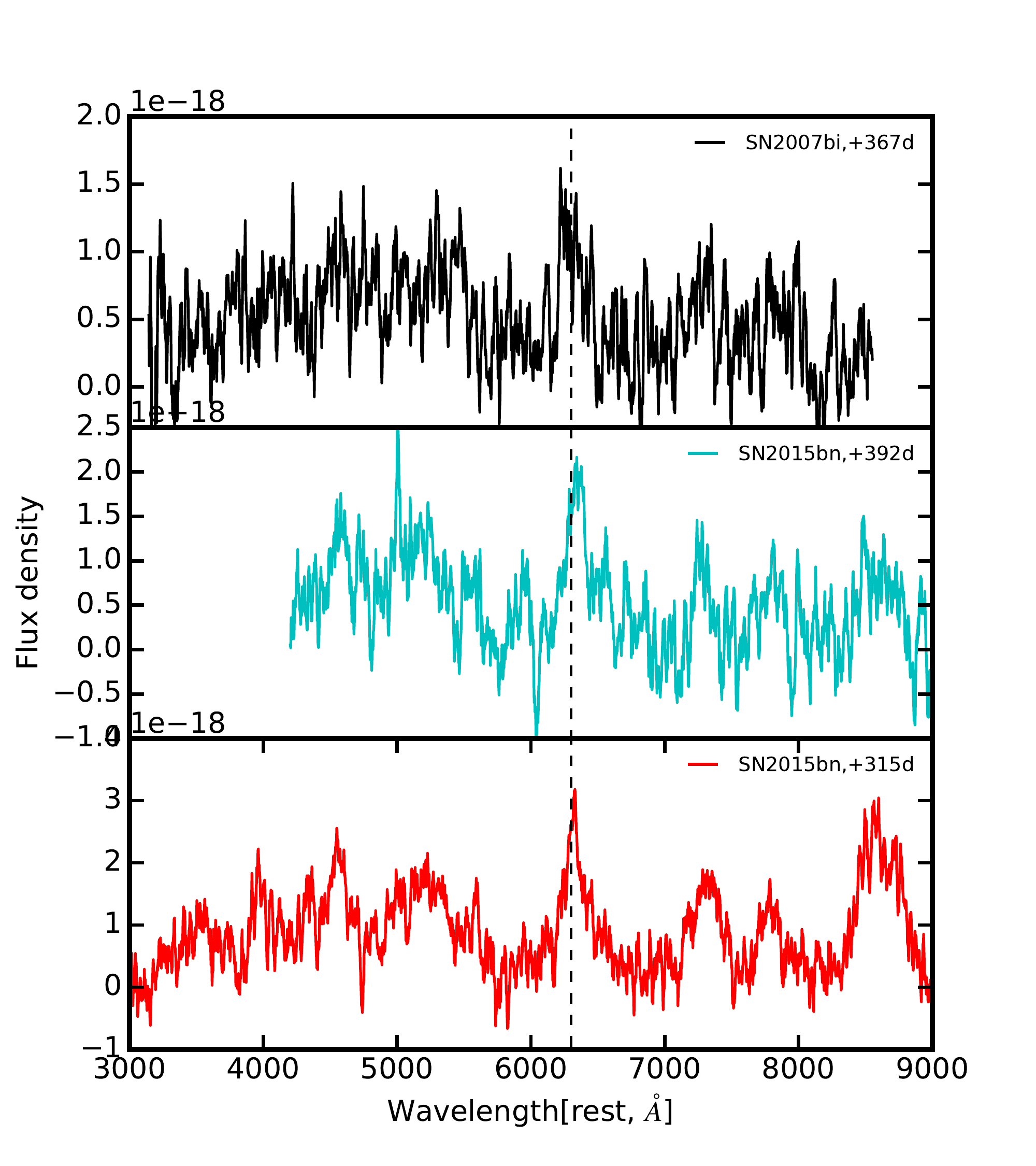}
}
\caption{ The simulated late-time spectra at the distance of iPTF16bad and with the noise of the $+241$\,day spectrum of iPTF16bad.  The input late-time spectra are from SN2007bi and SN2015bn. The dashed line marks the [O\,I]\,6300\,\AA\ emission. \label{simspec}}
\end{figure}

To address this question, we take three late-time spectra from SLSN-I SN2015bn at $+315$ and $+392$\,day \citep{Nicholl2016b}, SN2007bi at $+367$\,day \citep{Gal-Yam2009}, which all have prominent [O\,I]\,6300\,\AA\ emission.  We scale these three spectra to the distance of iPTF16bad, {\it i.e.} multiply by the square of the luminosity distance ratios, and then add the noise measured from the $+242$\,day spectrum for iPTF16bad.  The simulated spectra from this procedure are shown in Figure~\ref{simspec}.  Here the noise added to the input spectra is $\sigma \sim 9\times10^{-19}$\,erg\,s${-1}$\,sec$^{-2}$\,\AA$^{-1}$, measured from the 240\,\AA\ region centered at 6300\,\AA\ (excluding H$\alpha$) in the $+242$\,day spectrum of iPTF16bad.  This noise is then added as a Gaussian random noise to the input spectra. We note that because the input spectra do have their own noises,  the output simulated spectra may have slightly higher RMS than the true value.

From this simple simulation, we conclude that if the $+242$\,day spectrum from iPTF16bad were to have the same [O\,I] luminosity as that of the three input spectra, we would have detected this feature in our data.  
This implies that our spectra at $\sim +240$\,days likely have intrinsically weaker [O\,I]\,6300\,\AA\ emission than the late-time spectra of the three comparison SLSNe-I. Unfortunately, at $z\sim 0.2 - 0.3$, the Ca\,II triplet at 8498\AA\ is redshifted out of the optical range,  so we do not know if the Ca\,II triplet is strong, and serves as an alternative cooling line for these three events. In the simulated spectra, the Ca\,II triplet is quite strong in SN2015bn, and absent in SN2007bi  (see \citealt{Gal-Yam2009}).

The [O I] 6300, 6364 lines at nebular phase are very useful diagnostics of supernova ejecta. This is because these lines are  efficient coolant and typically re-emit a large fraction of heating energy of the O-material. More importantly, these lines become optically thin, and line luminosity $\propto M_{OI} n_e e^{-\Delta E/T(t)} $ in the Non-local thermal equilibrium (NLTE) phase, as discussed in details in \citep{Jerkstrand2017b}. 

The apparent weakness of the [O\,I]\,6300 doublet and other ionized O emissions in our three events seem to suggest that there is less oxygen emitting at late-times.  This is puzzling because SLSNe-I are thought to be explosive events of C+O cores with masses $>$ several tens of solar masses.  There could be several explanations. First, the ejecta of our three events may have lower
oxygen masses, and perhaps also lower progenitor masses than that of SN2015bn and SN2007bi. The recent modeling of the nebular spectra of SN2015bn and SN2007bi by \citet{Jerkstrand2016} has derived $M(O) \sim 10 - 30\,M_\odot$.  The second possible explanation is that these three events could be pulsational pair-instability supernova (PPISN).  Calculated optical spectra at nebular phase based on pair-instability supernova (PISN) models seem to show a relatively weak [O\,I]\,6300 doublet \citep{Jerkstrand2016}. At face value, this could be considered as supporting evidence for a PISN or PPISN model for these events. However, as shown by \citet{Jerkstrand2016},  the calculated nebular spectra show very strong [Ca\,II]\,7300\AA\ emission lines, which are also not detected in our late-time spectra, but are present in the simulated spectra (Figure\ref{simspec}).  The probability of these three events being PPISN or PISN is small. First is because the required progenitor mass is very high, and secondly, as pointed by \citet{Woosley2016}, PPISN still has difficulties producing very energetic SLSNe-I, and iPTF13ehe is such an example.

The third possible explanation is that Oxygen in these three events is detached from $^{56}$Ni, and with very little mixing.  If the O-zone is above $^{56}$Ni,  $\gamma$-ray photons from $^{56}$Ni decay will be effectively absorbed by Fe group elements before reaching O. In this case, there is not sufficient photon heating to produce [O\,I] emission.  If ejecta density is very high, it would result in high opacity, and the [O\,I] line may cool inefficiently.  Future better modeling of nebular spectra of SLSNe-I would narrow down these possible explanations.

\subsection{Nature of the LC undulations in iPTF15esb}
\label{subsec_15esbLC}

What makes iPTF15esb stand out is its peculiar LC with strong undulations, particularly in bluer bands. We note that the three peaks are separated from each other equally by $\sim22$\,days.  After the first peak, its $g$-band and also bolometric LC have two additional small bumps (Figure~\ref{obsLC} \&\ \ref{bolo}).

\begin{figure}
\includegraphics[width=0.5\textwidth,height=0.5\textwidth]{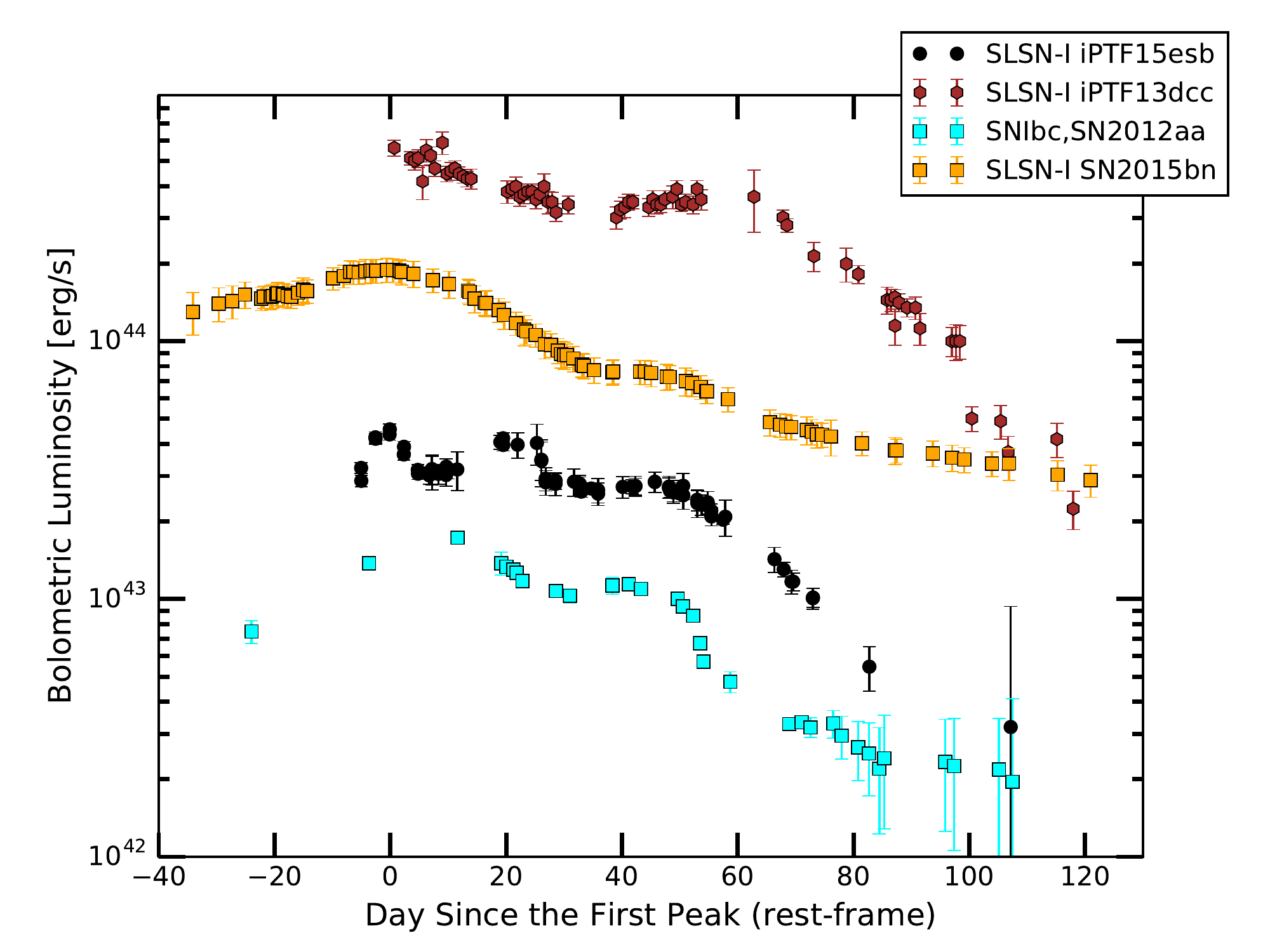}
\caption{This plot compares the bolometric light curves of iPTF15esb, SN2012aa and SLSNe-I SN2015bn and iPTF13dcc. SN2012aa is a SN between SN\,Ibc and SLSN-I \citep{Roy2016}. 
\label{LCcomp}}
\end{figure}

LC undulations are also seen in other SN types, such as SN2012aa (between SN\,Ibc and SLSN-I), other SLSNe-I (SN2015bn), SN\,IIn (PTF13z  and SN2009ip) \citep{Roy2016, Nicholl2016,Pastorello2013, Nyholm2017,Inserra2017}. They are probably even present in SN 2007bi and PS1-14bj \citep{Gal-Yam2009, Lunnan2016}.
SN2009ip is either a SN\,IIn or a SN imposter \citep{Fraser2013,Graham2014,Margutti2014,Martin2013}.  Figure~\ref{LCcomp} makes a LC comparison between iPTF15esb, SN2012aa and SN2015bn. The LC undulation in iPTF15esb is quite strong, with some similarity to that of SN2012aa.  These LC undulations are clearly very different from the double-peak LCs with initial weak bumps followed by prominent main peaks seen in LSQ14bdq, SN2006oz, PTF12dam and iPTF13dcc \citep{Leloudas2012, Nicholl2015,Smith2016, Vreeswijk2016}.  Their physical nature may also be different.

Several ideas were proposed by previous studies to explain the light curves. This includes (1) successive collisions between ejecta and mass shells expelled by previous episodic mass losses; (2)  magnetar UV-breakout predicted by \citet{Metzger2014}; (3) recombination of certain ionized elements; (4) variable continuum optical/UV opacities which could modulate the photon diffusion, thus affecting LC morphology, as proposed for ASASSN-15lh \citep{Godoy-Rivera2017, Margutti2017}.

In the case of iPTF15esb, interaction based models may explain the data.  The second peak has a duration of 20\,days and with a net luminosity of $1.5\times10^{43}$\,erg\,s$^{-1}$. The total extra energy in this peak is $\sim 2\times10^{49}$\,erg. 
This implies an excess luminosity of $\sim 10^{43}$\,erg\,s$^{-1}$ over 10\,days. 
Using a simple scaling relation $L \sim {1\over2}M_{csm} v^2/\delta t$ and taking the ejecta velocity $v_{ej}\sim 17,000$\,km\,s$^{-1}$, we estimate $M_{csm} \sim 0.01M_\odot$. Eruptive mass losses could produce such mass shells.  PPISN models \citep{Woosley2016} could produce various successive H-poor shells, and the subsequent collisions between shells and/or ejecta-shell would generate additional energy producing the observed LC undulations. 

The second possible explanation is due to the change in recombination of elements such as C and O, as shown in \citet{Piro2014},  CSM with pure CO can undergo recombination at temperatures of roughly 8000\,K.  Another similar idea is the change of continuum opacity, which can naturally explain the stronger undulation in bluer bands \citep{Godoy-Rivera2017, Margutti2017}. 
Finally, models with central power sources, such as magnetars or fall-back accretion onto a neutron star or black hole, 
have an energy input function, such as ${E_p \over \tau_p} {1 \over (1 + t/\tau_p)^2}$, with $E_p$ and $\tau_p$ as the magnetic dipole spin-down energy and the spin-down time scale respectively \citep{Dexter2013,Kasen2010}. At late-times, the luminosity should scale like t$^{-2}$, close to $(t-50)^{-2.5}$, measured from the data (Figure~\ref{bolo}).  These models seem to be able to explain some data. However, the real test requires detailed calculations which can meet the challenges of all observed features.   

\section{Summary and conclusions  }
\label{sec_summ}

We report two new SLSNe-I with broad H$\alpha$ emission in their late-time spectra discovered by iPTF. Together with iPTF13ehe \citep{Yan2015}, we now have three such events at $z\sim0.2 - 0.3$. 
The H$\alpha$ line luminosities reach as high as $(1-3)\times10^{41}$\,erg\,s$^{-1}$ and the line widths range from $4000 - 10,000$\,km\,s$^{-1}$.   We highlight four key observational results from our data.  

First, we interpret the late-time H$\alpha$ emission as a result of ejecta interaction with a neutral H-shell. The shock heating ionizes the neutral H atoms, which subsequently recombine and produce H$\alpha$ emission.
The detection of H$\alpha$ lines around $100-300$\,days since explosion imply that the H-shell must be at a distance not much farther than $10^{16}$\,cm from the progenitor star.  This shell with mass $\le 30\,M_\odot$ and an expansion speed of several 100\,km\,s$^{-1}$  indicates a very energetic mass loss, which must have happened not much longer than $10-30$ years prior to the supernova explosion. Such a tight timing provides a strong constraint on evolutionary models of massive stars.  

The short time interval inferred from our H-shell model also implies that progenitor stars can not have had time to also lose all of the helium envelope. Therefore, it is likely that the ejecta of these three SLSNe-I may have some helium.  The real mystery is why we do not detect any He features in the early-time spectra, even though at early times, our three events may have sufficient non-thermal energy sources to ionize He in the ejecta, as suggested by the detections of a partial O\,II absorption series.

Second, the $\sim 250$\,day spectra of two of our events show no detectable [O\,I]\,6300\,\AA\ emission. Using simulations, we demonstrate that at these late phases, our events have intrinsically lower [O\,I]\,6300\,\AA\ luminosities in comparison with $200 - 395$\,day spectra of SLSN-I SN2007bi and SN2015bn. Several different  scenarios could explain this observation. The simplest one is that the ejecta of our three events may have oxygen masses
less than $10 - 30\,M_\odot$, which was estimated for SN2015bn and SN2007bi \citep{Jerkstrand2017}.  

The third result is that for all three events, we initially see that the broad H$\alpha$ lines are blue-shifted relative to the hosts.  This may be in tension with the massive binary model because it predicted that we should see both red-shifted and blue-shifted H$\alpha$ emission lines relative to the host galaxies \citep{Moriya2015}. Interestingly, the velocity offsets between the broad H$\alpha$ and the host galaxies change from positive to negative with time for two of the events.  The very late-time spectra ($+125$ and $+242$\,days) of iPTF15esb and iPTF16bad show a weak signal of excess emission red-ward of 6563\,\AA.  We propose that a decrease of obscuration with time could be a possible explanation for seeing through more H$\alpha$ emission from the back side of the H-shells.

Finally,  the LC of iPTF15esb has a distinct morphology with significant undulations and three peaks are separated equally by $\sim22$\,days. Together with the evidence of H-shells,  these observations paint a picture of extended and multiple CSM shells, or CSM clumps,  at different radii.  
The LC undulation could be explained by H-poor-ejecta CSM interaction.  This would require some mechanisms which can eject multiple layers of material from massive progenitor stars within a time interval of several decades before supernova explosion. One possibility is Pulsational Pair-instability supernova \citep{Woosley2016}.  Quantitative modelings of our data are needed.
 
With these intriguing results, one important question is how representative these three objects are among other SLSNe-I.  Are they unique compared to other SLSNe-I?  Photometrically, these three events are very different from each other, in their peak luminosities, and post-peak decay rates and LC morphology. However, their LC properties fall within the diverse range shown by published SLSNe-I, and are not much different from the general population of SLSN-I.  Spectroscopically, these three events are very similar to each other at both early and late-times. However, they show some marked differences from other SLSNe-I, with their lack of full O\,II absorption series in the early-time spectra, and their higher ejecta velocities.
In conclusion, our three events are not completely peculiar, but probably represent a subset of the general SLSN-I population. At face value, these three events represent (10-15)\%\ of the PTF SLSN-I sample, although the number of events with good late-time follow-up is not large. The situation is definitely improving.

\acknowledgments

We thank N. Blagorodnova, A. Ho, Y. Cao, H. Vedantham, V. Ravi at Caltech, Virginia Cunningham at University of Maryland, Antonino Cucchiara at Goddard Space Flight Center (GSFC),  J. Mauerhan, I. Shivvers and P. Kelly at University of Berkeley for taking some of the data for iPTF15esb.  This paper benefitted from discussions with Anthony Piro at Carnegie Observatories.
The intermediate Palomar Transient Factory project is a scientific collaboration among the California Institute of Technology, Los Alamos National Laboratory, the University of  Wisconsin, Milwaukee,  the  Oskar  Klein  Center,  the Weizmann Institute of Science, the TANGO Program of the  University  System  of  Taiwan,  and  the  Kavli  Institute  for  the  Physics  and  Mathematics  of  the  Universe. LANL participation in iPTF is supported by the US Department of Energy as a part of the Laboratory Directed Research and Development program. A portion of this work was carried out at the Jet Propulsion Laboratory under a Research and Technology Development Grant, under contract with the National Aeronautics and Space Administration. This paper made use of Lowell Observatory's Discovery Channel Telescope (DCT). Lowell operates the DCT in partnership with Boston University, Northern Arizona University, the University of Maryland, and the University of Toledo. Partial support of the DCT was provided by Discovery
Communications. Large Monolithic Imager (LMI) on DCT was built by Lowell Observatory using funds from the National Science Foundation (AST-1005313). DAH, CM, and GH are funding by NSF grant AST-1313484.  This paper makes use of data from Las Cumbres Observatory.
This work was partially supported by the GROWTH project funded by the National Science Foundation under Grant No 1545949. This research has made use of the NASA/IPAC Extragalactic Database (NED) which is operated by the Jet Propulsion Laboratory, California Institute of Technology, under contract with the National Aeronautics and Space Administration. Some of the data presented herein were obtained at the W.M. Keck Observatory, which is operated as a scientific partnership among the California Institute of Technology, the University of California and the National Aeronautics and Space Administration. The Observatory was made possible by the generous financial support of the W.M. Keck Foundation. The authors wish to recognize and acknowledge the very significant cultural role and reverence that the summit of Mauna Kea has always had within the indigenous Hawaiian community.  We are most fortunate to have the opportunity to conduct observations from this mountain.

{\it Facilities:} \facility{Palomar}, \facility{Keck}, \facility{Discovery Channel Telescope}.

\clearpage
\begin{deluxetable*}{ccccccccccc}
\tablecolumns{7}
\tablewidth{0pc}
\tabletypesize{\scriptsize}
\tablecaption{\label{sample} Basic Properties of the SLSNe-I in the Sample}
\tablehead{\colhead{Name} & \colhead{RA}  & \colhead{DEC} & \colhead{Redshift} & \colhead{$E(B-V)$} & \colhead{$u_{host}$} & \colhead{$g_{host}$} & \colhead{$r_{host}$} & \colhead{$i_{host}$} & \colhead{$z_{host}$} \\
                                            & J2000 & J2000 &                         &             mag                       & mag           &  mag             &   mag         &       mag  & mag }
 \startdata
\input{table2.tex}
 \enddata
\tablenotetext{a}{iPTF13ehe is not within the SDSS foot print. Here \nodata means we did not obtain the host galaxy photometry in that band.}
\tablenotetext{b}{iPTF16bad is within the SDSS foot print, but not detected in all five bands. The magnitude limits quoted here are the 50\%\ completeness limits measured by the SDSS survey \citep{Abazajian2003}.}
\end{deluxetable*}

\begin{deluxetable*}{cccccc}
\tabletypesize{\footnotesize}
\tablewidth{0pt}
\tablecaption{The Spectroscopic Observation Log
\label{tab:spec}}
\tablehead{
\colhead{Object} &
\colhead{Obs.Date} &
\colhead{Phase$^a$} &
\colhead{Instrument} &
\colhead{Exp.Time$^b$} &
\colhead{Inst. Res.$^c$} \\
% units of columns
%\hline \\
 &  & days &  & seconds & \AA \\
}
\startdata
\input{table1.tex}
\tablenotetext{a}{The rest-frame phases for iPTF15esb and iPTF16bad are relative to the peak dates of MJD = 57363.5 and 57540.4\,day respectively.}
\tablenotetext{b}{Keck/LRIS exposure times for blue and red side can be different.}
\tablenotetext{c}{Instrument resolution is measured as Full-Width-Half-Maximum (FWHM) of unresolved sky lines.}
\tablenotetext{d}{The transient signals have mostly faded in these two spectra. They are dominated by the host galaxy light.}
\enddata
\end{deluxetable*}

\begin{deluxetable*}{ccccccccccccc}
\tablecolumns{7}
\tablewidth{0pc}
\tabletypesize{\scriptsize}
\tablecaption{\label{tab:phot} Photometry for iPTF15esb}
\tablehead{\colhead{Name} & \colhead{Filt}  & \colhead{MJD} & \colhead{Mag} & \colhead{Err} & \colhead{Filt} & \colhead{MJD} & \colhead{Mag} & \colhead{Err}  & \colhead{Filt} & \colhead{MJD} & \colhead{Mag} & \colhead{Err}  \\
                                           &                   & day            & mag                              &  mag    &                                &  day            &   mag                       &       mag    &                            &  day           &  mag         &  mag             }
 \startdata
\input{table3.tex}
 \enddata
\tablenotetext{a}{All magnitudes are in AB system. No extinction correction has been applied.}
\end{deluxetable*}

\begin{deluxetable*}{ccccccccccccc}
\tablecolumns{7}
\tablewidth{0pc}
\tabletypesize{\scriptsize}
\tablecaption{\label{tab:phot2} Photometry for iPTF16bad}
\tablehead{\colhead{Name} & \colhead{Filt}  & \colhead{MJD} & \colhead{Mag} & \colhead{Err} & \colhead{Filt} & \colhead{MJD} & \colhead{Mag} & \colhead{Err}  & \colhead{Filt} & \colhead{MJD} & \colhead{Mag} & \colhead{Err}  \\
                                           &                   & day            & mag                              &  mag    &                                &  day            &   mag                       &       mag    &                            &  day           &  mag         &  mag             }
 \startdata
\input{table3_part2.tex}
 \enddata
\tablenotetext{a}{All magnitudes are in AB system. No extinction correction has been applied.}
\end{deluxetable*}

\begin{deluxetable}{ccccccc}
\tablecolumns{7}
\tablewidth{0pc}
\tabletypesize{\scriptsize}
\tablecaption{\label{tab_linefit} Spectral Line Fit Results$^a$ }
\tablehead{\colhead{Name} & \colhead{Phase$^b$}  & \colhead{[OII]} & \colhead{N.H$\alpha$} & \colhead{B.H$\alpha$} & $V_{FWHM}$ & $\Delta V$ \\
& day & erg/s & erg/s & erg/s & km/s & km/s \\ }
 \startdata
\input{table4.tex}
 \enddata
 \tablenotetext{a}{Here N.H$\alpha$ and B.H$\alpha$ refer to the narrow and broad H$\alpha$ component respectively.}
 \tablenotetext{b}{The rest-frame days relative to the peak date.}
\end{deluxetable}

\end{document}

%% file: table2.tex
iPTF13ehe & 06:53:21.50 & +67:07:56.0 & 0.3434 & 0.04 & \nodata$^a$ & 24.9 & 24.24 & \nodata & \nodata \\
iPTF15esb & 07:58:50.67 & +66:07:39.1 & 0.224  & 0.04 & 23.65 & 22.61 & 21.90 & 21.50 & 21.44\\
iPTF16bad & 17:16:39.73 & +28:22:12.6  & 0.2467 & 0.04 & $>22.4^b$ & $>22.6$ & $>22.6$ & $>21.7$ & $>20.9$ \\

%% file: table1.tex
iPTF15esb  & 2015-12-07 & +0   & Keck/DEIMOS & 300 & 4 \\
iPTF15esb  & 2015-12-16 & +7.4   & Keck/DEIMOS & 600 & 4 \\
iPTF15esb  & 2016-01-07 & +25.3    & Keck/DEIMOS & 600 & 4 \\
iPTF15esb  & 2016-01-12 & +29.9    & Keck/LRIS   & 600(b),600(r) & 5.6 \\
iPTF15esb  & 2016-02-02 & +46.5    & P200/DBSP   & 1800 & 6 \\
iPTF15esb & 2016-02-07 & +50.9     & Keck/LRIS   & 1800(b),1800(r) & 5.6 \\
iPTF15esb & 2016-02-09 & +52.3   & Keck/LRIS   & 1200(b), 1200(r) & 5.6 \\
iPTF15esb & 2016-03-06 & +73.8   & Keck/LRIS   & 1200(b),1200(r) & 5.6 \\ 
iPTF15esb & 2016-04-10 & +102.4   & Keck/LRIS   & 1841(b),1800(r) & 5.6 \\
iPTF15esb & 2016-05-05 & +122.8   & Keck/LRIS   & 3000(b),2850(r) & 5.6 \\
iPTF15esb & 2016-11-02 & +270.9$^d$   & Keck/LRIS   & 1800(b),1720(r) & 5.6 \\
iPTF15esb & 2017-01-02 & +320.6$^d$   & Keck/LRIS   & 5400(b),5100(r) & 5.6 \\
iPTF16bad & 2016-06-04 & +2.5   & Keck/DEIMOS &  240             & 4 \\
iPTF16bad & 2016-06-07 & +5.3   & Keck/LRIS   & 240(b),240(r)   & 5.6 \\
iPTF16bad & 2016-09-30 & +97.3   & Keck/LRIS   & 2900(b),2700(r) & 5.6 \\
iPTF16bad & 2017-03-29 & +242.0   & Keck/LRIS   & 3040(b),2800(r)  & 5.6 \\

%% file: table3.tex
g & 57371.732 & 20.05 & 0.06 & r & 57371.733 & 19.90 & 0.06 & i & 57371.734 & 19.83 & 0.07 \\
g & 57373.663 & 20.07 & 0.08 & r & 57373.664 & 19.84 & 0.05 & i & 57373.665 & 20.00 & 0.06 \\
g & 57375.658 & 20.06 & 0.06 & r & 57375.659 & 19.84 & 0.06 & i & 57375.661 & 19.75 & 0.07 \\
g & 57377.652 & 20.09 & 0.18 & r & 57377.653 & 19.69 & 0.08 & i & 57377.654 & 19.95 & 0.11 \\
g & 57386.748 & 19.71 & 0.05 & r & 57386.749 & 19.55 & 0.04 & i & 57386.750 & 19.70 & 0.05 \\
g & 57403.879 & 20.52 & 0.06 & r & 57403.880 & 19.83 & 0.05 & i & 57403.882 & 19.90 & 0.07 \\
g & 57405.993 & 20.60 & 0.07 & r & 57405.995 & 19.84 & 0.06 & i & 57405.996 & 19.87 & 0.07 \\
g & 57412.618 & 20.63 & 0.14 & r & 57408.604 & 19.74 & 0.15 & i & 57408.606 & 19.87 & 0.11 \\
g & 57414.632 & 20.61 & 0.09 & r & 57412.619 & 19.86 & 0.08 & i & 57410.711 & 20.11 & 0.16 \\
g & 57422.622 & 20.61 & 0.08 & r & 57414.634 & 19.79 & 0.06 & i & 57412.620 & 19.96 & 0.08 \\
g & 57424.605 & 20.57 & 0.08 & r & 57416.650 & 19.80 & 0.04 & i & 57414.635 & 19.8 & 0.07 \\
g & 57430.603 & 20.63 & 0.13 & r & 57422.623 & 19.82 & 0.06 & i & 57416.651 & 19.85 & 0.06 \\
g & 57444.706 & 21.51 & 0.18 & r & 57430.604 & 19.84 & 0.06 & i & 57422.625 & 19.91 & 0.04 \\
g & 57446.656 & 21.70 & 0.09 & r & 57444.707 & 20.56 & 0.08 & i & 57424.608 & 19.74 & 0.09 \\
g & 57448.671 & 22.00 & 0.14 & r & 57446.657 & 20.63 & 0.06 & i & 57430.606 & 20.17 & 0.11 \\
g & 57452.814 & 22.11 & 0.16 & r & 57448.672 & 20.69 & 0.06 & i & 57444.709 & 20.33 & 0.1 \\
g & 57452.854 & 22.06 & 0.13 & r & 57452.846 & 20.80 & 0.05 & i & 57446.658 & 20.59 & 0.1 \\
g & 57403.800 & 20.39 & 0.04 & r & 57464.623 & 21.09 & 0.10 & i & 57448.673 & 20.58 & 0.09 \\
g & 57433.830 & 20.82 & 0.03 & r & 57472.637 & 21.83 & 0.18 & i & 57452.811 & 20.33 & 0.14 \\
g & 57464.770 & 22.33 & 0.06 & r & 57472.667 & 21.74 & 0.11 & i & 57452.850 & 20.45 & 0.08 \\
g & 57494.650 & 22.59 & 0.08 & r & 57473.711 & 21.71 & 0.10 & i & 57458.699 & 20.84 & 0.11 \\
g & 57357.326 & 20.00 & 0.07 & r & 57473.768 & 21.85 & 0.17 & i & 57463.622 & 20.91 & 0.12 \\
g & 57357.357 & 20.12 & 0.09 & r & 57475.723 & 21.93 & 0.19 & i & 57464.6258 & 20.9 & 0.13 \\
g & 57360.337 & 19.70 & 0.06 & r & 57475.729 & 21.90 & 0.21 & i & 57472.670 & 21.18 & 0.11 \\
g & 57360.378 & 19.71 & 0.04 & r & 57479.655 & 21.74 & 0.20 & i & 57480.733 & 21.88 & 0.19 \\
g & 57363.340 & 19.67 & 0.05 & r & 57480.727 & 22.16 & 0.16 & i & 57483.713 & 21.92 & 0.18 \\
g & 57363.380 & 19.62 & 0.07 & r & 57483.707 & 22.12 & 0.15 & i & 57403.800 & 19.70 & 0.03 \\
g & 57366.347 & 19.87 & 0.06 & r & 57403.800 & 19.72 & 0.03 & i & 57433.830 & 19.85 & 0.03 \\
g & 57366.387 & 19.79 & 0.07 & r & 57433.830 & 19.98 & 0.03 & i & 57464.77 & 20.77 & 0.02 \\
g & 57369.306 & 20.02 & 0.08 & r & 57464.770 & 21.08 & 0.02 & i & 57494.65 & 21.32 & 0.04 \\
g & 57369.343 & 20.05 & 0.06 & r & 57494.650 & 21.70 & 0.05 & i & 57396.851 & 19.80 & 0.17 \\
g & 57372.293 & 19.91 & 0.11 & r & 57394.988 & 19.31 & 0.17 & i & 57396.855 & 20.06 & 0.19 \\
g & 57372.329 & 19.98 & 0.13 & r & 57395.931 & 19.70 & 0.10 & i & 57398.934 & 19.84 & 0.07 \\
g & 57375.287 & 20.16 & 0.10 & r & 57395.935 & 19.61 & 0.13 & i & 57398.937 & 19.74 & 0.05 \\
g & 57375.327 & 19.94 & 0.06 & r & 57396.844 & 20.0 & 0.09 & i & 57404.0468 & 19.97 & 0.06 \\
g & 57387.335 & 19.63 & 0.04 & r & 57396.847 & 19.74 & 0.06 & i & 57404.0493 & 19.9 & 0.05 \\
g & 57387.375 & 19.80 & 0.05 & r & 57398.929 & 19.69 & 0.04 & i & 57407.932 & 19.91 & 0.08 \\
g & 57390.412 & 19.88 & 0.12 & r & 57398.932 & 19.69 & 0.04 & i & 57407.934 & 19.84 & 0.08 \\
g & 57402.318 & 20.755 & 0.20 & r & 57404.042 & 19.83 & 0.04 & i & 57411.988 & 19.62 & 0.12 \\
g & 57422.377 & 20.50 & 0.12 & r & 57404.044 & 19.76 & 0.03 & i & 57411.990 & 19.69 & 0.09 \\
g & 57422.407 & 20.47 & 0.12 & r & 57407.927 & 19.89 & 0.07 & i & 57415.829 & 19.8 & 0.09 \\
g & 57425.399 & 20.34 & 0.13 & r & 57407.930 & 19.81 & 0.05 & i & 57415.831 & 19.86 & 0.08 \\
g & 57425.429 & 20.66 & 0.17 & r & 57411.983 & 19.66 & 0.11 & i & 57419.928 & 19.79 & 0.06 \\
g & 57428.371 & 20.67 & 0.13 & r & 57411.985 & 19.74 & 0.16 & i & 57419.931 & 19.78 & 0.07 \\
g & 57428.402 & 20.80 & 0.18 & r & 57415.824 & 19.89 & 0.03 & i & 57423.827 & 19.95 & 0.07 \\
g & 57431.337 & 20.89 & 0.09 & r & 57415.826 & 19.79 & 0.03 & i & 57423.829 & 19.84 & 0.07 \\
g & 57431.368 & 21.14 & 0.14 & r & 57419.923 & 19.78 & 0.04 & i & 57429.820 & 19.94 & 0.05 \\
g & 57434.313 & 20.89 & 0.23 & r & 57419.926 & 19.82 & 0.04 & i & 57429.823 & 19.94 & 0.05 \\
g & 57394.483 & 19.88 & 0.20 & r & 57423.822 & 19.81 & 0.04 & i & 57442.913 & 20.16 & 0.11 \\                
g & 57395.421 & 19.90 & 0.20 & r & 57423.824 & 19.96 & 0.04 & i & 57442.917 & 20.14 & 0.14 \\
g & 57395.426 & 19.87 & 0.16 & r & 57429.812 & 19.96 & 0.03 & i & 57448.823 & 20.3 & 0.08 \\
g & 57396.334 & 20.08 & 0.10 & r & 57429.816 & 20.01 & 0.02 & i & 57448.827 & 20.32 & 0.07 \\
g & 57396.339 & 19.99 & 0.09 & r & 57436.952 & 20.18 & 0.15 & i & 57455.749 & 20.68 & 0.10 \\
g & 57398.422 & 20.22 & 0.1 & r & 57442.909 & 20.41 & 0.13 & i & 57455.753 & 20.56 & 0.12 \\
g & 57398.426 & 20.13 & 0.07 & r & 57448.816 & 20.62 & 0.05 & i & 57455.838 & 20.32 & 0.11 \\
g & 57403.534 & 20.48 & 0.1 & r & 57448.819 & 20.66 & 0.05 & i & 57455.841 & 20.76 & 0.17 \\
g & 57403.538 & 20.5 & 0.09 & r & 57455.830 & 20.92 & 0.10 &  \nodata & \nodata & \nodata & \nodata \\
g & 57407.419 & 20.52 & 0.14 & r & 57455.834 & 20.92 & 0.11 & \nodata & \nodata & \nodata & \nodata \\
g & 57407.423 & 20.65 & 0.14 & r & 57462.894 & 21.26 & 0.15 & \nodata & \nodata & \nodata & \nodata \\
g & 57415.316 & 20.44 & 0.09 & r & 57463.850 & 21.47 & 0.14 & \nodata & \nodata & \nodata & \nodata \\
g & 57415.320 & 20.43 & 0.11 & r & 57463.854 & 21.23 & 0.19 & \nodata & \nodata & \nodata & \nodata \\
g & 57419.416 & 20.26 & 0.1 & \nodata & \nodata & \nodata & \nodata & \nodata & \nodata & \nodata & \nodata \\
g & 57419.419 & 20.25 & 0.1 & \nodata & \nodata & \nodata & \nodata & \nodata & \nodata & \nodata & \nodata \\
g & 57423.314 & 20.48 & 0.11 & \nodata & \nodata & \nodata & \nodata & \nodata & \nodata & \nodata & \nodata \\
g & 57423.318 & 20.42 & 0.11 & \nodata & \nodata & \nodata & \nodata & \nodata & \nodata & \nodata & \nodata \\
g & 57429.302 & 20.61 & 0.11 & \nodata & \nodata & \nodata & \nodata & \nodata & \nodata & \nodata & \nodata \\
g & 57429.307 & 20.62 & 0.11 & \nodata & \nodata & \nodata & \nodata & \nodata & \nodata & \nodata & \nodata \\
g & 57448.311 & 21.56 & 0.13 & \nodata & \nodata & \nodata & \nodata & \nodata & \nodata & \nodata & \nodata \\

%% file: table3_part2.tex
g & 57539.921 & 19.10 & 0.06 & r & 57539.964 & 19.46 & 0.08 & i & 57547.058 & 19.87 & 0.08 \\
g & 57546.978 & 19.20 & 0.02 & r & 57547.021 & 19.46 & 0.04 & i & 57547.060 & 19.81 & 0.07 \\
g & 57547.908 & 19.41 & 0.02 & r & 57547.903 & 19.41 & 0.02 & i & 57547.905 & 19.8 & 0.05 \\
g & 57551.980 & 19.43 & 0.05 & r & 57550.961 & 19.56 & 0.07 & i & 57551.977 & 19.88 & 0.08 \\
g & 57555.085 & 19.63 & 0.23 & r & 57551.974 & 19.56 & 0.03 & i & 57574.014 & 20.42 & 0.15 \\
g & 57555.089 & 19.57 & 0.05 & r & 57556.050 & 19.76 & 0.12 & i & 57574.018 & 20.53 & 0.16 \\
g & 57556.042 & 19.82 & 0.21 & r & 57556.052 & 19.78 & 0.09 & i & 57575.957 & 20.71 & 0.13 \\
g & 57556.046 & 19.77 & 0.10 & r & 57563.049 & 19.84 & 0.14 & i & 57575.960 & 20.69 & 0.12 \\
g & 57563.045 & 20.01 & 0.16 & r & 57563.051 & 19.71 & 0.09 & i & 57579.783 & 20.57 & 0.07 \\
g & 57574.002 & 20.84 & 0.12 & r & 57573.779 & 20.31 & 0.04 & i & 57581.009 & 20.58 & 0.21 \\
g & 57575.954 & 20.81 & 0.08 & r & 57574.007 & 20.25 & 0.15 & i & 57589.740 & 20.99 & 0.18 \\
g & 57575.958 & 20.78 & 0.07 & r & 57574.011 & 20.33 & 0.06 & i & 57592.783 & 20.96 & 0.07 \\
g & 57578.008 & 21.04 & 0.11 & r & 57575.959 & 20.37 & 0.06 & i & 57596.859 & 21.23 & 0.19 \\
g & 57578.013 & 20.98 & 0.17 & r & 57579.781 & 20.65 & 0.06 & i & 57598.781 & 21.01 & 0.09 \\
g & 57579.785 & 21.23 & 0.06 & r & 57580.998 & 20.61 & 0.08 & i & 57601.872 & 21.2 & 0.12 \\
g & 57580.988 & 21.23 & 0.12 & r & 57581.001 & 20.66 & 0.11 & i & 57630.662 & 21.87 & 0.17 \\
g & 57580.993 & 21.10 & 0.10 & r & 57586.977 & 20.85 & 0.18 & \nodata & \nodata & \nodata & \nodata \\
g & 57592.789 & 21.94 & 0.10 & r & 57592.777 & 21.03 & 0.05 & \nodata & \nodata & \nodata & \nodata \\
g & 57596.846 & 21.91 & 0.19 & r & 57596.851 & 21.32 & 0.1 & \nodata & \nodata & \nodata & \nodata \\
g & 57598.787 & 22.10 & 0.15 & r & 57596.855 & 21.08 & 0.1 & \nodata & \nodata & \nodata & \nodata \\
g & 57601.855 & 22.40 & 0.22 & r & 57598.775 & 21.38 & 0.12 & \nodata & \nodata & \nodata & \nodata \\
\nodata & \nodata & \nodata & \nodata & r & 57600.920 & 21.33 & 0.23 & \nodata & \nodata & \nodata & \nodata \\
\nodata & \nodata & \nodata & \nodata & r & 57601.865 & 21.41 & 0.10 & \nodata & \nodata & \nodata & \nodata \\
\nodata & \nodata & \nodata & \nodata & r & 57601.869 & 21.33 & 0.13 & \nodata & \nodata & \nodata & \nodata \\
\nodata & \nodata & \nodata & \nodata & r & 57618.741 & 21.85 & 0.21 & \nodata & \nodata & \nodata & \nodata \\
\nodata & \nodata & \nodata & \nodata & r & 57630.659 & 22.17 & 0.22 & \nodata & \nodata & \nodata & \nodata \\

%% file: table4.tex
15esb & 30  & 8.8e+39   & 5.3e+39  & \nodata  & \nodata  & \nodata \\
15esb & 50  & 4.9e+39   & 7.4e+39  & \nodata  & \nodata  & \nodata \\
15esb & 52  & 1.0e+40   & 7.4e+39  & \nodata  & \nodata  & \nodata \\
15esb & 73  & 8.9e+39   & 5.6e+39  & 2.6e+41  & 10800    & 1051    \\
15esb & 102 & 1.0e+40   & 7.5e+39  & 8.4e+40  & 5382     & -160   \\
15esb & 122 & 8.1e+39   & 1.0e+39  & 5.4e+40   & 3768     & -69   \\
15esb & 270  & 1.6e+40  & 8.5e+39  & \nodata   & \nodata  & \nodata  \\
16bad & 125  & 4.8e+39  & 3.9e+39  & 1.0e+41   & 5930     & 366    \\
16bad & 242  & 2.2e+39  & 3.6e+39  & 2.2e+40   & 3225     & -549   \\ 
13ehe & +251 & \nodata  & 6.1e+39  & 1.5e+41   & 4852     & 457 \\
13ehe & +254 & \nodata  & 6.2e+39  & 1.04e+41  & 4312     & 457 \\
13ehe & +278 & \nodata  & 1.24e+40 & 9.9e+40   & 4096     & 366 \\

%% file: ms.bbl
\begin{thebibliography}{}
\bibitem[Abazajian et al.(2003)]{Abazajian2003} Abazajian, K., Adelman-McCarthy, J.~K., Ag{\"u}eros, M.~A., et al.\ 2003, \aj, 126, 2081 


\bibitem[Arnett(1996)]{Arnett96} Arnett, D.\ 1996, Supernovae
and Nucleosynthesis: An Investigation of the History of Matter, from the
Big Bang to the Present, by D.~Arnett.~Princeton: Princeton University
Press, 1996.,
\bibitem[Benetti et al.(2014)]{Benetti2014} Benetti, S., Nicholl, M., Cappellaro, E., et al.\ 2014, \mnras, 441, 289 

\bibitem[Blanton et al.(2003)]{Blanton2003} Blanton, M.~R., Hogg, D.~W., Bahcall, N.~A., et al.\ 2003, \apj, 592, 819 

\bibitem[Cardelli et al.(1989)]{Cardelli1989} Cardelli, J.~A., Clayton, G.~C., \& Mathis, J.~S.\ 1989, \apj, 345, 245 
\bibitem[Chugai(1991)]{Chugai1991} Chugai, N.~N.\ 1991, \mnras, 250, 513   
\bibitem[Dilday et al.(2012)]{Dilday2012} Dilday, B., Howell, D.~A., Cenko, S.~B., et al.\ 2012, Science, 337, 942 

      
\bibitem[Dexter \& Kasen(2013)]{Dexter2013} Dexter, J., \& Kasen, D.\ 2013, \apj, 772, 30

\bibitem[Faber et al.(2003)]{Faber2003} Faber, S.~M., Phillips, 
A.~C., Kibrick, R.~I., et al.\ 2003, \procspie, 4841, 1657 
\bibitem[Filippenko et al.(1993)]{Filippenko1993} Filippenko, A.~V., Matheson, T., \& Ho, L.~C.\ 1993, \apjl, 415, L103 


\bibitem[Fox et al.(2015)]{Fox2015} Fox, O.~D., Silverman, J.~M., Filippenko, A.~V., et al.\ 2015, \mnras, 447, 772 
\bibitem[Fransson et al.(2002)]{Fransson2002} Fransson, C., Chevalier, R.~A., Filippenko, A.~V., et al.\ 2002, \apj, 572, 350 

\bibitem[Fraser et al.(2013)]{Fraser2013} Fraser, M., Inserra, C., Jerkstrand, A., et al.\ 2013, \mnras, 433, 1312 
\bibitem[Fremling et al.(2016)]{Fremling2016} Fremling, C., Sollerman, J., Taddia, F., et al.\ 2016, \aap, 593, A68 

\bibitem[Gal-Yam (2017)]{Gal-Yam2017} Gal-Yam, A. \ 2017, \apj, in prep.

\bibitem[Gal-Yam et al.(2009)]{Gal-Yam2009} Gal-Yam, A., Mazzali, P., Ofek, E.~O., et al.\ 2009, \nat, 462, 624 


\bibitem[Gal-Yam(2012)]{Gal-Yam2012} Gal-Yam, A.\ 2012, Science, 337, 927
\bibitem[Gal-Yam et al.(2014)]{Gal-Yam2014} Gal-Yam, A., Arcavi, I., Ofek, E.~O., et al.\ 2014, \nat, 509, 471 


\bibitem[Gezari et al.(2009)]{Gezari2009} Gezari, S., Halpern, J.~P., Grupe, D., et al.\ 2009, \apj, 690, 1313 


\bibitem[Graham et al.(2014)]{Graham2014} Graham, M.~L., Sand, D.~J., Valenti, S., et al.\ 2014, \apj, 787, 163 
\bibitem[Georgy et al.(2012)]{Georgy2012} Georgy, C., Ekstr{\"o}m, S., Meynet, G., et al.\ 2012, \aap, 542, A29 
\bibitem[Godoy-Rivera et al.(2017)]{Godoy-Rivera2017} Godoy-Rivera, D., Stanek, K.~Z., Kochanek, C.~S., et al.\ 2017, \mnras, 466, 1428 


\bibitem[Hachinger et al.(2012)]{Hachinger2012} Hachinger, S., Mazzali, P.~A., Taubenberger, S., et al.\ 2012, \mnras, 422, 70 



\bibitem[Jerkstrand et al.(2016)]{Jerkstrand2016} Jerkstrand, A., Smartt, S.~J., \& Heger, A.\ 2016, \mnras, 455, 3207 
\bibitem[Jerkstrand et al.(2017)]{Jerkstrand2017} Jerkstrand, A., Smartt, S.~J., Inserra, C., et al.\ 2017, \apj, 835, 13 
\bibitem[Jerkstrand(2017)]{Jerkstrand2017b} Jerkstrand, A.\ 2017, arXiv:1702.06702 



\bibitem[Kasen \& Bildsten(2010)]{Kasen2010} Kasen, D., \& Bildsten, L.\ 2010, \apj, 717, 245 

\bibitem[Inserra et al.(2016a)]{Inserra2016a} Inserra, C., Smartt, S.~J., Gall, E.~E.~E., et al.\ 2016, arXiv:1604.01226 
\bibitem[Inserra et al.(2016)]{Inserra2016} Inserra, C., Fraser, M., Smartt, S.~J., et al.\ 2016, \mnras, 459, 2721 
\bibitem[Inserra et al.(2014)]{Inserra2014} Inserra, C., Smartt, S.~J., Scalzo, R., et al.\ 2014, \mnras, 437, L51 
\bibitem[Inserra et al.(2013)]{Inserra2013} Inserra, C., Smartt, S.~J., Jerkstrand, A., et al.\ 2013, \apj, 770, 128 
\bibitem[Inserra et al.(2017)]{Inserra2017} Inserra, C., Nicholl, M., Chen, T.-W., et al.\ 2017, arXiv:1701.00941 
\bibitem[Liu et al.(2016)]{Liu2016} Liu, Y.-Q., Modjaz, M., \& Bianco, F.~B.\ 2016, arXiv:1612.07321 



\bibitem[Langer(2012)]{Langer2012} Langer, N.\ 2012, \araa, 50, 107 


\bibitem[Leloudas et al.(2012)]{Leloudas2012} Leloudas, G., Chatzopoulos, E., Dilday, B., et al.\ 2012, \aap, 541, A129
\bibitem[Leloudas et al.(2015)]{Leloudas2015} Leloudas, G., Schulze, S., Kr{\"u}hler, T., et al.\ 2015, \mnras, 449, 917 


\bibitem[Lunnan et al.(2016)]{Lunnan2016} Lunnan, R., Chornock, R., Berger, E., et al.\ 2016, \apj, 831, 144 
\bibitem[Lunnan et al.(2014)]{Lunnan2014} Lunnan, R., Chornock, R., Berger, E., et al.\ 2014, \apj, 787, 138 

 
\bibitem[Nicholl et al.(2016)]{Nicholl2016} Nicholl, M., Berger, E., Smartt, S.~J., et al.\ 2016, \apj, 826, 39
\bibitem[Nicholl et al.(2016b)]{Nicholl2016b} Nicholl, M., Berger, E., Margutti, R., et al.\ 2016, \apjl, 828, L18 
\bibitem[Nicholl et al.(2015)]{Nicholl2015} Nicholl, M., Smartt, S.~J., Jerkstrand, A., et al.\ 2015, \apjl, 807, L18 
\bibitem[Nicholl et al.(2014)]{Nicholl2014} Nicholl, M., Smartt, S.~J., Jerkstrand, A., et al.\ 2014, \mnras, 444, 2096 

\bibitem[Moriya et al.(2015)]{Moriya2015} Moriya, T.~J., Liu, Z.-W., Mackey, J., Chen, T.-W., \& Langer, N.\ 2015, \aap, 584, L5 


\bibitem[Masci et al.(2016)]{Masci2016} Masci, F., Laher, R., Rebbapragada, U., et al.\ 2016, arXiv:1608.01733  
\bibitem[Margutti et al.(2014)]{Margutti2014} Margutti, R., Milisavljevic, D., Soderberg, A.~M., et al.\ 2014, \apj, 780, 21
\bibitem[Margutti et al.(2017)]{Margutti2017a} Margutti, R., Kamble, A., Milisavljevic, D., et al.\ 2017, \apj, 835, 140 

 
\bibitem[Margutti et al.(2017)]{Margutti2017} Margutti, R., Metzger, B.~D., Chornock, R., et al.\ 2017, \apj, 836, 25 
\bibitem[Martin(2013)]{Martin2013} Martin, J.\ 2013, Journal of the American Association of Variable Star Observers (JAAVSO), 41, 391 



\bibitem[Matheson et al.(2000)]{Matheson2000} Matheson, T., Filippenko, A.~V., Ho, L.~C., Barth, A.~J., \& Leonard, D.~C.\ 2000, \aj, 120, 1499 


\bibitem[Mazzali et al.(2016)]{Mazzali2016} Mazzali, P.~A., Sullivan, M., Pian, E., Greiner, J., \& Kann, D.~A.\ 2016, \mnras, 458, 3455 


  \bibitem[Metzger et al.(2014)]{Metzger2014} Metzger, B.~D., Vurm, I., Hasco{\"e}t, R., \& Beloborodov, A.~M.\ 2014, \mnras, 437, 703 
  \bibitem[Milisavljevic et al.(2015)]{Milisavljevic2015} Milisavljevic, D., Margutti, R., Kamble, A., et al.\ 2015, \apj, 815, 120 


  \bibitem[Miller et al.(2009)]{Miller2009} Miller, A.~A., Chornock, R., Perley, D.~A., et al.\ 2009, \apj, 690, 1303 
  \bibitem[Nyholm et al.(2017)]{Nyholm2017} Nyholm, A., Sollerman, J., Taddia, F., et al.\ 2017, arXiv:1703.09679
%\bibitem[Nyholm et al.(2017)]{Nyholm2017} Nyholm, A. et al. \aa, in press

\bibitem[Ofek et al.(2007)]{Ofek2007} Ofek, E.~O., Cameron, P.~B., Kasliwal, M.~M., et al.\ 2007, \apjl, 659, L13 
\bibitem[Ofek et al.(2014)]{Ofek2014} Ofek, E.~O., Sullivan, M., Shaviv, N.~J., et al.\ 2014, \apj, 789, 104 


      
\bibitem[Oke \& Gunn(1982)]{Oke1982} Oke, J.~B., \& Gunn, J.~E.\ 1982, \pasp, 94, 586            
\bibitem[Oke et al.(1995)]{Oke1995} Oke, J.~B., Cohen, J.~G., Carr, M., et al.\ 1995, \pasp, 107, 375            
\bibitem[Piro \& Morozova(2014)]{Piro2014} Piro, A.~L., \& Morozova, V.~S.\ 2014, \apjl, 792, L11
\bibitem[Padmanabhan(2000)]{padmanabhan2000} Padmanabhan, T.\ 2000,
Theoretical Astrophysics - Volume 1, Astrophysical Processes, by
T.~Padmanabhan, pp.~622.~Cambridge University Press, December
2000.~ISBN-10: 0521562406.~ISBN-13: 9780521562409.~LCCN: QB461 .P33 2000,
\bibitem[Pastorello et al.(2013)]{Pastorello2013} Pastorello, A., Cappellaro, E., Inserra, C., et al.\ 2013, \apj, 767, 1 
\bibitem[Pastorello et al.(2010)]{Pastorello2010} Pastorello, A., Smartt, S.~J., Botticella, M.~T., et al.\ 2010, \apjl, 724, L16 

\bibitem[Perley et al.(2016)]{Perley2016} Perley, D.~A., Quimby, R.~M., Yan, L., et al.\ 2016, \apj, 830, 13 

         
\bibitem[Phinney(1989)]{Phinney1989} Phinney, E.~S.\ 1989, The Center of the Galaxy, 136, 543 
\bibitem[Planck Collaboration et al.(2016)]{Planck2015} Planck Collaboration, Ade, P.~A.~R., Aghanim, N., et al.\ 2016, \aap, 594, A13
  
%\bibitem[Planck Collaboration et al.(2015)]{Planck2015} Planck
%Collaboration, Ade, P.~A.~R., Aghanim, N., et al.\ 2015, arXiv:1502.01589

     
\bibitem[Quimby et al.(2011)]{Quimby2011} Quimby, R.~M., Kulkarni, 
S.~R., Kasliwal, M.~M., et al.\ 2011, \nat, 474, 487 
\bibitem[Quimby et al.(2017)]{Quimby2017} Quimby, R.~M., et al. \ 2017, \apj, in prep.

\bibitem[Roy et al.(2016)]{Roy2016} Roy, R., Sollerman, J., Silverman, J.~M., et al.\ 2016, \aap, 596, A67 


\bibitem[Schlafly \& Finkbeiner(2011)]{Schlafly2011} Schlafly, E.~F., \& Finkbeiner, D.~P.\ 2011, \apj, 737, 103     
\bibitem[Silverman et al.(2013)]{Silverman2013} Silverman, J.~M., Nugent, P.~E., Gal-Yam, A., et al.\ 2013, \apj, 772, 125 
\bibitem[Smith et al.(2016)]{Smith2016} Smith, M., Sullivan, M., D'Andrea, C.~B., et al.\ 2016, \apjl, 818, L8
  \bibitem[Smith et al.(2009)]{Smith2009} Smith, N., Silverman, J.~M., Chornock, R., et al.\ 2009, \apj, 695, 1334 
\bibitem[Smith et al.(2007)]{Smith2007} Smith, N., Li, W., Foley, R.~J., et al.\ 2007, \apj, 666, 1116 
\bibitem[Smith(2014)]{Smith2014} Smith, N.\ 2014, \araa, 52, 487 

\bibitem[Taddia et al.(2016)]{Taddia2016} Taddia, F., Fremling, C., Sollerman, J., et al.\ 2016, \aap, 592, A89 

\bibitem[Thomas(2013)]{Thomas2013} Thomas, R.~C.\ 2013, Astrophysics Source Code Library, ascl:1308.008 


\bibitem[Turatto et al.(1993)]{Turatto1993} Turatto, M., Cappellaro, E., Danziger, I.~J., et al.\ 1993, \mnras, 262, 128        

\bibitem[Vreeswijk et al.(2017)]{Vreeswijk2016} Vreeswijk, P.~M., Leloudas, G., Gal-Yam, A., et al.\ 2017, \apj, 835, 58
\bibitem[Woosley(2017)]{Woosley2016} Woosley, S.~E.\ 2017, \apj, 836, 244


%\bibitem[Vreeswijk et al.(2016)]{Vreeswijk2016} Vreeswijk, P.~M., Leloudas, G., Gal-Yam, A., et al.\ 2016, arXiv:1609.08145
%\bibitem[Woosley(2016)]{Woosley2016} Woosley, S.~E.\ 2016, arXiv:1608.08939
%\bibitem[Woosley \& Heger(2007)]{Woosley2007} Woosley, S.~E., \& Heger, A.\ 2007, \physrep, 442, 269 

\bibitem[Woosley et al.(2007)]{Woosley2007} Woosley, S.~E., Blinnikov, S., \& Heger, A.\ 2007, \nat, 450, 390 

\bibitem[Woosley et al.(2002)]{Woosley2002} Woosley, S.~E., Heger, A., \& Weaver, T.~A.\ 2002, Reviews of Modern Physics, 74, 1015 


\bibitem[Yan et al.(2015)]{Yan2015} Yan, L., Quimby, R., Ofek, E., et al.\ 2015, \apj, 814, 108
\bibitem[Yan et al.(2016)]{Yan2016} Yan, L., Quimby, R., Gal-Yam, A., et al.\ 2016, arXiv:1611.02782 
\bibitem[Yaron et al.(2017)]{Yaron2017} Yaron, O., Perley, D.~A., Gal-Yam, A., et al.\ 2017, arXiv:1701.02596 

\bibitem[Yaron \& Gal-Yam(2012)]{Yaron2012} Yaron, O., \& Gal-Yam, A.\ 2012, \pasp, 124, 668 



\end{thebibliography}
